\documentclass[12 pt]{article}

\usepackage[utf8]{inputenc}
\usepackage[T1]{fontenc}
\usepackage{natbib}
\usepackage{aas_macros}
\usepackage{graphicx}
\usepackage{amsmath}
\usepackage{siunitx}
\usepackage{footnote}
\usepackage{longtable}
\usepackage{tablefootnote}
\usepackage{authblk}
\usepackage{pdflscape}
\usepackage{hyperref}
\usepackage{caption}
\usepackage{subcaption}

\DeclareSIUnit\angstrom{\text {Å}}

    \makeatletter
\def\@fnsymbol#1{\ensuremath{\ifcase#1\or *\or  
   \mathsection\or \mathparagraph\or \|\or **\or \dagger\dagger
   \or \ddagger\ddagger \else\@ctrerr\fi}}
    \makeatother
  
\begin{document}
\title{Spectral Evolution and Photo-ionization Analysis of Nova Cas 2020 (V1391 Cas)}
%\date{November 2020}

\author[1]{G. M. Hamed\thanks{Email: gamal.hamed@nriag.sci.eg},\thanks{Visiting Astronomer, Istanbul University, Department of Astronomy \& Space Sciences} }
\author[2]{ H. H. Esenoglu}
\author[3]{A. I. Galeev\thanks{Passed away in February 2021}}
\affil[1]{Stellar Astronomy Lab, Astronomy Department, National Research Institute of Astronomy and Geophysics, Cairo, Egypt, 11421}
\affil[2]{Istanbul University, Department of Astronomy and Space Sciences, 34119, Beyazit, Istanbul, T\"{u}rkiye}
\affil[3]{Kazan Federal University, ul. Kremlevskaya 18, 420008, Kazan, Russia; Academy of Sciences of Tatarstan, Kazan, Russia}
\maketitle

\begin{abstract}
    We present spectroscopic observations of Nova Cas 2020(V1391 Cas) obtained using the Russian Turkish Telescope during different stages of its 2020 outburst. We followed the spectral evolution of the nova until it entered the nebular phase. The expansion velocity of the ejecta reached $\sim$780 $\mathrm{km\,s^{-1}}$. The fluxes of the neutral [O I] lines at wavelengths 6300, 6364, and 5577 $\si{\angstrom}$ were used to calculate the electron temperature and the mass of neutral oxygen in the ejecta. We found average values $\mathbf{T_e = 4890\,K}$ ,$\mathbf{M_{OI} = 2.54 \times 10^{-5}} M\textsubscript{\(\odot\)}$ which are consistent with the values calculated for other novae. We modeled the nova's ejected envelope 515 days after its discovery and found that the log elemental abundances by number relative to Hydrogen of the envelope are He =  -0.7, C = -5.5, O = -2.5, N = -2.0 and Ne = -4.0.    
\end{abstract}

%\begin{keywords}

%\end{keywords}

\section{Introduction}
The modern understanding of classical novae is that they are sudden outbursts appearing on cataclysmic explosions resulting from the thermonuclear runaway of the matter accreted on the surface of the white dwarf \citep{2016PASP..128e1001S,2020A&ARv..28....3D}. Nova Cas 2020 (V1391 Cas) was discovered by S. Korotkiy as the optical transient TCP J00114297+6611190 on 2020-07-27.9302 UT at an unfiltered magnitude of 12.9 and it was identified as a Fe II classical nova after its discovery \citep{2020ATel13903....1S}. To follow the spectral evolution of the nova, we took this discovery date as the start epoch ("\textbf{Day 0}") and provided it throughout the paper along with the observational dates. The nova showed multiple brightness enhancements during the maximum phase, followed by a dust formation period. Therefore, we summarized the spectroscopic evolution of the nova in chronological order with comments by reviewing all the literature including mostly the Astronomer's Telegrams in this section. The nova showed multiple secondary maxima for about three months after the visual maximum. It is the second nova to show both CN and $\mathrm{C_2}$ bands near the maximum (the first being V2676 Oph in 2012) indicating that the envelope is Carbon rich \citep{2021ApJ...907...70F}. The discoverers reported that the object was not visible down to magnitude 13.5 on images taken on 2020-07-06.9054 UT (\textbf{Day -21}) \citep{2020AAN...715....1W}. Pre-discovery images of the nova provide a constraint on the outburst date between \textbf{Day -1} and \textbf{Day 0} \citep{2020ATel14004....1S}. Determining the maximum of the nova outburst is crucial, as the system brightens by 6-19 magnitudes from its pre-nova state \citep{2003cvs..book.....W}.

The spectra used in this paper were obtained as a part of an observational campaign to observe and study the envelopes of classical novae in different stages of the outburst. This ongoing campaign started in mid 2020 and about 15 novae have been observed so far. This paper is the first of a series of papers about the results of this campaign. We observed V1391 Casusing the 1.5 m Russian Turkish Telescope (RTT150) to monitor its early spectroscopic evolution. Here we present our observations and analysis of V1391 Cas.

\subsection{Evolution of V1391 Cas by outburst date}
\textbf{Day 2}\\
The optical spectrum was dominated by H I, Fe II, and Na I P-Cygni profiles on the nova's spectrum observed on 2020-07-29.025 (\textbf{Day 2}). The absorption of the H$\alpha$ P-Cygni profile gave a velocity of around $-850\pm 100$ km s$^{-1}$ \citep{2020ATel13903....1S}.

The nova reached its visual maximum at 10.6 mag in the V band on 2020-07-29.917 UT (Day 2) \citep[e.g.][]{2021BAV.....1....1W}. After a short while, the spectrum of V1391 Cas taken by \citet{2020ATel13905....1M} on 2020-07-29.960 UT (\textbf{Day 2}), showed that the spectrum changed significantly from that of \citep{2020ATel13903....1S}. Accordingly, all P-Cyg lines were in emission, except for the minimal photospheric absorption lines and O I (7772 \AA), where absorption was -385 km $s^{-1}$. A red continuum was dominant. All emission lines appeared rather sharp, with an average FWHM of about 500 km $s^{-1}$. The emission in Na I and the P-Cyg absorption in H$\alpha$ (with integrated absolute flux of $3.543 \times 10^{-12}\, erg\, cm^{-2}s^{-1}$) disappeared, and He I $\mathbf{5876\AA}$ emerged in emission with a flux of $2.154 \times 10^{-14}\, erg\, cm^{-2}\,s^{-1}$ . The large interstellar extinction was quite evident in the significant strength of Diffuse Interstellar Bands (DIBs). The equivalent widths (EWs in $\si{\angstrom}$) of these DIBs were calculated as follows: 5780 (0.57), 5793 (0.3), and 6613 $\si{\angstrom}$ (0.37). The integrated absolute fluxes (in $10^{-12}$ erg cm$^{2}$ s$^{-1}$) of some representative emission lines observed in the spectrum were given as the following: H$\alpha$ (3.54), O I 7772 (0.70), H$\beta$ (0.25), H$\gamma$ (0.06), Fe II 42 5169 (0.06), H$\delta$ (0.02), He I 5876 (0.02), Fe II 38 4583 (0.01), N II 3 5682 (0.01).\\

\leftline{\textbf{Days 3-54}}
\citet{2020ATel13939....1S}, reported that the nova monitoring, which continued for about 12 days between 2020-07-30.7 (\textbf{Day 3}) and 2020-08-11.9 UT (\textbf{Day 16}), showed a very significant evolution in the spectra. Accordingly, the O I 7773 $\si{\angstrom}$ multiplet showed P-Cygni profiles in the first spectrum on 2020-07-30.8 (\textbf{Day 3}).

According to \citet{2021ApJ...907...70F}, optical low-resolution spectroscopic observations from 2020-07-31 to 2020-08-19 (\textbf{Days 4-23}) revealed the appearance and disappearance of molecular absorption bands of $\mathrm{C_2}$ and CN in V1391 Cas. These molecules were only present for $\sim$3 days. $\mathrm{C_2}$ and probably CN formed at $\sim$5000 K in the nova envelope. The spectral evolution and the formation conditions of the molecules are similar to those of V2676 Oph, which is the first example of a $\mathrm{C_2}$/CN-forming nova. They predicted that a late-phase grain formation episode similar to that seen in V2676 Oph will occur in V1391 Cas.

From 2020-08-1.8 (\textbf{Day 5}), Fe II , 5169, and other optical Fe II transitions showed P-Cygni profiles.

Meanwhile, \citet{2020ATel13919....1S} reported that an observation with Swift/XRT on 2020-08-02.888 UT (\textbf{Day 6}) showed no X-ray source at the position of the nova. Assuming power-law emission with the photon index of 2 and H I column density of 8.2x10$^{21}$ cm$^{-2}$ this translated to the unabsorbed 0.3-10 keV flux limit of 7x10$^{-14}$ erg cm $^{-2}$ s$^{-1}$.

Continuing the report of \citet{2020ATel13939....1S}, several blueshifted low-ionization metal absorbers, e.g. Mg II 4481, Ti II 4764, and Si II 6347, 6371 $\si{\angstrom}$, were present on 2020-08-4 (\textbf{Day 8}) and 2020-08-6 (\textbf{Day 10}) and persisted but were not well resolved in subsequent spectra.

On 2020-08-4.9 (\textbf{Day 8}), blueshifted absorption was seen in the Na I D1, D2 spectral lines at -190 km s$^{-1}$ with equivalent width EW = 0.1 $\si{\angstrom}$. The absorption radial velocity was the same as all other P-Cygni absorption on all Fe II lines (especially , 5169 $\si{\angstrom}$). 

The Si II doublet (6347, 6371 $\si{\angstrom}$) showed weak emission through 2020-08-6 (\textbf{Day 10}). There is a complex spectrum of weaker metal lines, all at low velocity, and the development of this nova shows similarities to nova DN Gem 1912 \citep{1965POMic...9..113M}. No evidence of He I 5876, 6678, and 7065 $\si{\angstrom}$ was found in any spectrum. 

The absorption profiles of Na I D1, D2 lines on 2020-08-6.8 (\textbf{Day 10}) increased to EW = 0.5 $\si{\angstrom}$ with no change in radial velocity. The metal line absorptions displayed consistently lower radial velocity than the Balmer lines, whose asymmetric absorptions were deepest (H$\alpha$, -250 km s$^{-1}$; H$\beta$, -210) while extending to about -620 km s$^{-1}$; the emission to absorption equivalent width ratios were 3.8 (H$\alpha$), 1.1 (H$\beta$). 

On 2020-08-8.3 (\textbf{Day 12}), the Ca II IR transitions showed weak emission, while O I 7773, 8446 $\si{\angstrom}$ both showed P-Cyg profiles (absorption depths of about 60\% and 25\%, respectively), similar to the Pa series. The 7773 $\si{\angstrom}$ multiplet absorption was complex and broad, centered at -180 km s$^{-1}$, but extending to +300 km s$^{-1}$ (in red), with the emission extending to about -500 km s$^{-1}$ (in blue). 

From the date of first detection, 2020-07-27.23087 (\textbf{Day 0}), the nova showed a series of flares (each lasting days to a week) with the brightest flare peaking at V=10.8 on 2020-08-10.08738 (\textbf{Day 14}) \citep{2020ATel14004....1S}. On 2020-08-10.72 (\textbf{Day 14}), as reported by \citet{2020ATel13939....1S}, the spectrum was dominated by the continuum and \citet{2021ApJ...907...70F} identified blueshifted absorption lines of neutral and low-ionized atomic species, such as CI, NI, OI, SiII, and FeII.

\citet{2020ATel13941....1F} reported that the molecular absorption bands of C$_2$ and CN were not evident in the spectra taken on 2020-08-10.692 UT (\textbf{Day 14}).

\citet{2020ATel13998....1H} reported in the optical spectroscopic observations on 2020-08-10.9 (\textbf{Day 14}) (The flux here is in units of $\mathbf{\times 10^{-14}\, erg cm^{-2}}$ $\mathbf{s^{-1}}$ and EW in $\mathbf{\si{\angstrom}}$.) that most of the Balmer series lines including H$\delta$ (flux=14.4, EW=2.9), H$\gamma$ (flux=13, EW=3), and H$\beta$ (flux=11, EW=1.9) were in absorption except for H$\alpha$ (absorption: flux=12.2, EW=0.9; emission: flux=52, EW=3.9) which exhibited a prominent P-Cygni profile. Also, other absorption lines were as follows: CaII H 3934 (flux=10.4, EW=3.2) and CaII K 3969 (flux=13.3, EW=4.1), HeI 4921 (flux=0.8, EW=0.1). The CaII doublet here was strong and blue-shifted. The He I 5876, He II 4686, and N II 5682 lines were absent in these spectra. The O I 7772 line showed a P-Cygni profile (absorption: flux=216, EW=3.9; emission: flux=48.1, EW=0.9), and the Na I 5896 interstellar absorption line also showed a strong component. 

Spectra on 2020-08-11.1 (near the light curve peak, \textbf{Day 15}) showed narrow P Cygni profiles of Balmer, O I, Fe II, and Na I \citep{2020ATel14004....1S}. The absorption troughs were at blueshifted velocities of 200-250 km s$^{-1}$, while the emission features of P Cygni profiles were relatively weak compared to absorption. This dramatic apparent deceleration of the absorption features (e.g. the absorption in the H$\alpha$ P-Cygni profile on \textbf{Day 2} was $-850\pm 100 \mathrm{km s^{-1}}$) was probably due to the photosphere receding to inner (slower moving) regions of the expanding ejecta.

In their observations on 2020-08-11.7 (\textbf{Day 15}), \citet{2020ATel13998....1H} also reported that most of the Balmer series lines including H$\delta$ (flux=1.3, EW=0.8), H$\gamma$ (flux=6.7, EW=3.1), and H$\beta$ (flux=9, EW=2.3 ) remained in prominent absorption except for H$\alpha$ (absorption: flux=6.8, EW=0.5; emission: flux=42.6, EW=3.4) which exhibited the P-Cygni profile. Soon after the EW of the absorption component of H$\alpha$ with the P-Cyg profile increased to about 3 $\si{\angstrom}$ and the EW of the emission component was 3.5 $\si{\angstrom}$. The He II 4686 and N II 5682 lines were also absent in these spectra. The O I 7772 line showed a P-Cygni profile (absorption: flux=31.7, EW=3.5; emission: flux=8.6, EW=0.9). The interstellar Na I 5896 and HeI 5876 (flux=17.6, EW=3.7) absorption lines also showed a strong component. The Ca II doublet 3934 (flux=7, EW=5.5) and 3968 (flux=6.4, EW=4.7) were strong and blueshifted. The blue continuum of the spectra decreased slightly from 2022-08-10.9 (\textbf{Day 14}) to 2020-08-11.7 (\textbf{Day 15}) by about one order of magnitude.

On 2020-08-11.9 (\textbf{Day 15}), the flux upper limit for the He I 5876 Å emission was <1.2x10$^{-15}$ erg cm$^{-2}$ s$^{-1}$. Based on the velocity range of the interstellar Na I D components ((V(LSR)=-46 to -9 km s$^{-1}$) and the width of the profile (140 km s$^{-1}$), the LAB survey yields a total N(H)=8.2x10$^{21}$ cm$^{-2}$ \citep{2005A&A...440..775K}. With an average velocity of -65 km s$^{-1}$, H I displayed a higher velocity than the Na I components, so this value is a possible upper limit for the interstellar extinction in the line of sight (the profile of Na I vs H I 21 cm also suggests that the nova is located at an intermediate distance from the edge of the Galaxy. In some cases it would be possible to provide a range for the distance of the nova based on the measured velocity of Na I interstellar lines and the profile of the H I line to the direction of the nova). In light of the still-rising optical flux and the lack of any forbidden lines (including [O I]), the nova may be still in the optically thick stage with the initiation of the recombination wave known from other novae in the pre-maximum stage.

For their observations on 2020-08-12.14 (\textbf{Day 16}), \citet{2020ATel13998....1H} reported that similar to their previous observations most of the Balmer series lines remained in prominent absorption except for H$\delta$, H$\gamma$ (flux=4, EW=6) H$\beta$ (flux=18.4, EW=11 $\si{\angstrom}$), and H$\alpha$ (absorption: EW=2.2, emission: EW=4.5) which exhibited a P Cygni profile. The He I 4921 absorption line (flux=3.1, EW=0.7) in the first spectra transformed to emission (flux=2.2, EW=1.2) in the later spectra. Similarly, in the later spectra, the absorption of H$\alpha$ in P-Cygni disappeared and the emission component (flux=164, EW=22.4) was very strong. The spectra also showed a strong emission line at a wavelength of 3451 $\si{\angstrom}$ with an EW of 6 $\si{\angstrom}$. The He I 5876 line was visible. However, the He II 4686 and N II 5682 lines were still absent. The O I 7772 line (absorption: flux=17.9, EW=1.6; emission: flux=13.6, EW=1.2) showed a P-Cygni profile and the Na I 5896 interstellar absorption line also showed a strong component. The Ca II doublet 3934 (flux=1.4, EW=3.6) and 3968 in absorption were strong and blueshifted. Here, too, the flux is in units of x10$^{-14}$ erg cm$^{-2}$ s$^{-1}$ and EW in $\si{\angstrom}$.

The low-resolution (R$\sim$500) spectrum taken by \citet{2020ATel13941....1F} on 2020-08-12.706 UT (\textbf{Day 16}) also first detected the molecular absorption bands of C$_2$ and CN a few days after the visual brightness maximum of the nova \citep{2021ApJ...907...70F}. For typical novae ($\sim$8000K; \citealp{2005MNRAS.360.1483E}) the effective temperatures of the so-called photosphere would be expected to be too high to form a simple function of the form $\nu^{\beta} B(\nu, T)$, where B is the Planck function at temperature T and $\beta (\simeq1)$ is a constant (the so-called $\beta$ index for the dust), to infrared spectra in the range 2–24 $\mu$m, and concluded that the grains eventually grew to $\sim$0.7 $\mu$m. V1391 Cas was colder than the typical novae on the date. As seen in the novae DQ Her and V2676 Oph, the C$_2$ and CN molecular band features that form when the brightness is near their maximum can disappear within about one week \citep{2017gacv.workE..64K}. These molecular absorption bands of C$_2$ and CN were also present on 2020-08-13.37 (\textbf{Day 17}) but not significant on 2020-08-14.79 (\textbf{Day 18}). They had almost vanished on 2020-08-15.69 (\textbf{Day 19}) \citep{2021ApJ...907...70F}. Finally, the observations of \citet{2020ATel13941....1F} on 2020-08-17.53 (\textbf{Day 21}) confirmed that the C$_2$ and CN absorption lines disappeared.

\citet{2021OEJV..220...37D} made simultaneous photometric and spectroscopic observations of V1391 Cas from 2020-08-14 to 22 (\textbf{Day 18-26}), 2020-08-28 to 09-22 (\textbf{Day 32-57}), and 2020-10-06 to 12-09 (\textbf{Day 74-135}). The evolution of H$\alpha$'s flux, equivalent width, and FWHM values from low-resolution spectra (R$\sim$1000) \textbf{concerning} the light curve in the V band of the nova were given.

The spectra between 2020-08-14 (\textbf{Day 18}) and 2020-08-16 (\textbf{Day 20}) showed broad emission lines with FWZI of about 2500 km s$^{-1}$ while the slow velocity absorption features of about 250 km s$^{-1}$ were superimposed on top of the broad emission \citep{2020ATel14004....1S}.

After the maximum (10.8 V-mag), the nova became fainter and fell to $\sim$13 V-mag on 2020-08-17 (\textbf{Day 21}) \citep{2021ApJ...907...70F}.

The detection of these C$_2$ and CN molecular absorption bands prompted \citet{2020ATel13967....1R} to search for carbon monoxide in the infrared. They found strong emissions from both the CO fundamental and the first overtone in infrared spectroscopic measurements of Nova V1391 Cas obtained on 2020-08-17.54 (\textbf{Day 21}). The atomic emission lines are dominated by C I. While other low excitation features were also presented (H I, N I, O I, Na I, Ca II, and Fe II), the C I multiplet at 1.0691 microns was the strongest feature in the spectral range. Numerous other C I transitions were also available. At the time of these measurements, there was no indication of the presence of dust in the material ejected from the nova, but dust formation was predicted to almost certainly occur in the next few weeks.

The spectrum obtained on 2020-08-20 (\textbf{Day 24}), which coincided with one of the secondary maxima ($\sim$11.5 V-mag, \citealp{2021ApJ...907...70F}), showed strong absorption features with absorption troughs at a velocity of around -400 km s$^{-1}$. In some lines, these absorptions were superimposed on top of the broad emission (FWZI $\sim$2800 km s$^{-1}$) \citep{2020ATel14004....1S}.

The spectra from 2020-08-22 to 2020-08-25 (\textbf{Days 26}-\textbf{29}) showed weakening absorption \citep{2020ATel14004....1S}.

\citet{2020ATel14006....1B} reported near-infrared spectroscopic measurements of V1391 Cas on 2020-09-02.614 UT (\textbf{Day 37}). Accordingly, the spectrum was typical of Fe II class novae, rich in CI lines while also displaying other usual lines of HI, OI, NI, and NaI. HeI 1.083 microns and 2.059 microns were present but weak. P-Cygni features were absent or very weakly seen in some of the lines (e.g. OI 1.3164 micron). The observed FWHM and FWZI values of the Paschen beta 1.2818-micron line, which showed a double-peaked structure, were 770 and 1930 km s$^{-1}$, respectively. The first overtone CO emission (\textbf{Day 22}) reported in \citet{2020ATel13967....1R} still persisted with the $^{12}$CO and $^{13}$CO band heads well resolved. There was still no definitive evidence for dust between \textbf{Day 22} and \textbf{Day 38}. Spectra on 2020-09-04 (\textbf{Day 39}), coinciding with another flare peak, showed strong absorption with absorption troughs at blueshifted velocities of 600-700 km s$^{-1}$, especially in H$\beta$ and Fe II; H$\alpha$ also showed absorption at -400 km s$^{-1}$ \citep{2020ATel14004....1S}. A similar behavior was observed in other novae with multiple maxima, where absorption lines strengthen around the peak brightness and appear at higher velocities (e.g. \citet{2011PASJ..63...911T}). Such spectral evolution might be understood if the flares were associated with multiple ejections each causing the photosphere to move outwards (with a new bunch of fast-moving material) during the peak and then to recede.

\citet{2020ATel14034....1W} reported infrared spectroscopic measurements of V1391 Cas on 2020-09-19.297 UT (\textbf{Day 54}). Accordingly, the spectra of the nova showed that all the lines (\textbf{Day 38}) mentioned by \citet{2020ATel14006....1B} were present and no significant evolution was seen in them. However, the first overtone CO emission on the \textbf{Day 22} \citep{2020ATel13967....1R} appeared to have either disappeared or it could be present at a very weak level.\\

\leftline{\textbf{Days 139-141}}
\citet{2020ATel14267....1M} suggested that the formation of dust had already started based on the optical spectrum of V1391 Cas taken at 2020-12-13.811 UT (\textbf{Day 139}). Here the brightness of the nova had dropped sharply. The spectrum had a very red slope, and the only emission lines (non-P-Cyg absorption), all showing a narrow profile, and together with their fluxes (x10$^{-14}$ erg cm$^{-2}$ s$^{-1}$) were as follows: H$\beta$ (3), FeII 4923 (1.2), FeII  (1.1), FeII 5169 (1), [OI] 6300 (6.7), [OI] 6364 (2.2), H$\alpha$ (80.7), and OI 7772 (9.5).

Follow-up observations in the near IR showed that the nova formed optically thick dust about 141 days (2020-12-15.73) after its discovery \citep{2020ATel14272....1B}. They gave the reddening as E(B-V)$\sim$1.4 mag. The (J-K) color was in the range of 3.2-3.55, indicating that the nova generated a significant amount of dust. They gave the dust temperature between 1200-1300 K.

The progenitor of V1391 Cas, in the GAIA DR3 catalog has a source ID of 528178486608462720 (distance=5.7 kpc and $T_{eff}$=7068 K) with RA=$00^{h}$ $11^{m}$ $42^{s}$.96 and DEC=+$66^{o}$ $11^{'}$ $20^{''}$.80 (alerting magnitude G=11.65; \footnote{\url{http://gsaweb.ast.cam.ac.uk/alerts/alert/Gaia20eld}}).

\citet{2020BAVSR..69..193K} calculated the distance of the nova to be 4.1 kpc. Here, the maximum luminosity of V1391 Cas was taken as 10.7 mag, the absolute luminosity as -6.5 mag, and the interstellar extinction value as A=4.2 mag. According to the light curve ($t_3>100$ days), V1391 Cas was a slow nova \citep[e.g.][]{2021BAV.....1....1W}. -6 to -10 mag is the range of variation for all novae (fast and slow). Also, according to \citet{1997PASP..109.1285E}, this value is $\leq-7.1$ mag which was found with a new approach to the nova velocity classes (see also Dissertation: \footnote{\url{https://tez.yok.gov.tr/UlusalTezMerkezi/tezSorguSonucYeni.jsp}}).

\section{Spectroscopic Observations}
The observations were made using the TUG Faint Object Spectrograph and Camera (TFOSC) mounted to the focal plane on the Russian Turkish Telescope (RTT150) of the TUBITAK National Observatory (TUG) in Bakirlitepe, Antalya, T\"{u}rkiye. Detailed information about the specifications of the telescope and its attached instruments can be found here \footnote{\url{https://tug.tubitak.gov.tr/en/teleskoplar/rtt150-telescope-0}} and here \footnote{\url{https://tug.tubitak.gov.tr/tr/icerik/tfosc-tug-faint-object-spectrograph-and-camera}}. Three grisms covering the spectral ranges 4220-6650$\si{\angstrom}$ \, 6190-8190$\si{\angstrom}$ \, and 3650-8740$\si{\angstrom}$ \,with spectral resolutions ($\lambda / \bigtriangleup \lambda$) of 1331 ($\bigtriangleup \lambda = 4.1$), 2189 ($\bigtriangleup \lambda = 3.0$), and 749 ($\bigtriangleup \lambda = 12.0$, broad-band) respectively, were used. We used an Andor iKon-L 936 BEX2-DD-9ZQ CCD camera. The observations were made on 7 nights in 2020 and 2021. A total of 17 spectra are presented. Table \ref{tab:log} shows a log of the spectra used in this work. 

\begin{landscape}

\begin{table}
\caption{Observation log of RTT150 TFOSC spectra.}
\label{tab:log}
\begin{footnotesize}

%\resizebox{1.2\columnwidth}{!}{%
\begin{longtable*}{ccccccccc}

\hline \hline

  DATE   &     TIME   &   JD           &   Days               &  Exposure  & Standard & $\lambda$ Range  & \textit{R} & Airmass \\%& Average S/N    \\
         &     (UT)   &   (2459000+)   &   (after Discovery)* &  (s)       & Star & ($\si{\angstrom}$)  & (Grism) &  \\
  \hline
 2020-08-10 & 21:19:19.60 &   072.38842  &13.96 (Day 14)    &  1200 & BD+28D4211    &      3650-8740 &  749 (15)   &         1.34  \\%&  153.9  \\   
%            & 21:46:33.70 &   072.40733  &13.98    &  600  &               &                &        &         1.294181  \\%&  117.0  \\   
            & 22:29:39.50 &   072.43726  &14.01 (Day 14)    &  600  &               &      4200-6650 &  1331 (7)  &         1.23  \\%&  97.67  \\   
            & 23:07:26.20 &   072.4635   &14.03 (Day 14)    &  600  &               &      6190-8190 &  2189 (8)  &         1.19  \\%&  126.7  \\   
 2020-08-11 & 19:16:31.00 &   073.30314  &14.87 (Day 14)    &  600  &   BD+28D4211   &      3650-8740 &  749 (15)   &         1.70  \\%&  91.64  \\   
            & 21:57:17.70 &   073.41479  &14.98 (Day 14)    &  600  &               &      4200-6650 &  1331 (7)  &         1.27  \\%&  38.07  \\   
 %           & 22:09:33.00 &   073.4233   &14.99    &  900  &               &                &        &         1.252409  \\%&  41.76  \\   
            & 22:28:05.00 &   073.43617  &15.01 (Day 15)    & 900  &               &      6190-8190 &  2189 (8)  &         1.23  \\%&  79.92  \\   
 2020-08-12 & 01:43:37.10 &   074.57196  &16.14 (Day 16)    &  600  & GD 248        &      3650-8740 &  749 (15)   &         1.16  \\%&  79.77  \\    
            & 01:55:47.20 &   074.58041  &16.15 (Day 16)    &  600  &               &      6190-8190 &  2189 (8)  &         1.17  \\%&  65.35  \\   
            & 02:07:47.30 &   074.58874  &16.16 (Day 16)    &  600  &               &      4200-6650 &  1331 (7)  &         1.18  \\%&  67.12  \\   
 2020-09-08 & 02:21:05.66 &   101.59798  &43.17 (Day 43)    &  900  & B0+25d4655    &      4200-6650 &   1331 (7)     &         1.35  \\%&  49.42  \\   
            & 02:37:36.36 &   101.60945  &43.18 (Day 43)    &  900  &               &      6190-8190 &  2189 (8)  &         1.37  \\%&  89.83  \\   
 2020-11-15 & 00:46:13.80 &   169.5321   &111.10 (Day 111)   &  600  & G191B2B       &      4200-6650 &  1331 (7)  &         1.91  \\%&  48.63  \\   
            & 00:58:56.30 &   169.5321  &111.11 (Day 111)   &  900  &               &      6190-8190 &  2189 (8)  &         1.98  \\%&  76.93  \\   
 2020-12-17 & 18:58:19.10 &   201.2905   &142.86 (Day 143)   &  1200 & G191B2B      &      4200-6650 &  1331 (7)  &         1.25  \\%&  22.22  \\   
            & 19:54:19.80 &   201.2905   &142.90 (Day 143)   &  1800 &               &      6190-8190 &  2189 (8)  &         1.34  \\%&  14.54  \\   
 2021-12-24 & 17:22:08.40 & 573.2237 & 514.79 (Day 515) & 3600 & G191B2B & 4200-6650 &  1331 (7)  & 1.18 \\
            & 18:26:16.80 & 573.2683 & 514.84 (Day 515) & 3600 &       & 6190-8190 & 2189 (8) & 1.24 \\
\hline\\
{* Day 0 is JD 2459058.43}
\end{longtable*}%
%}
\end{footnotesize}

\end{table}
\end{landscape}

\begin{figure}[ht]
\includegraphics[width=\textwidth]{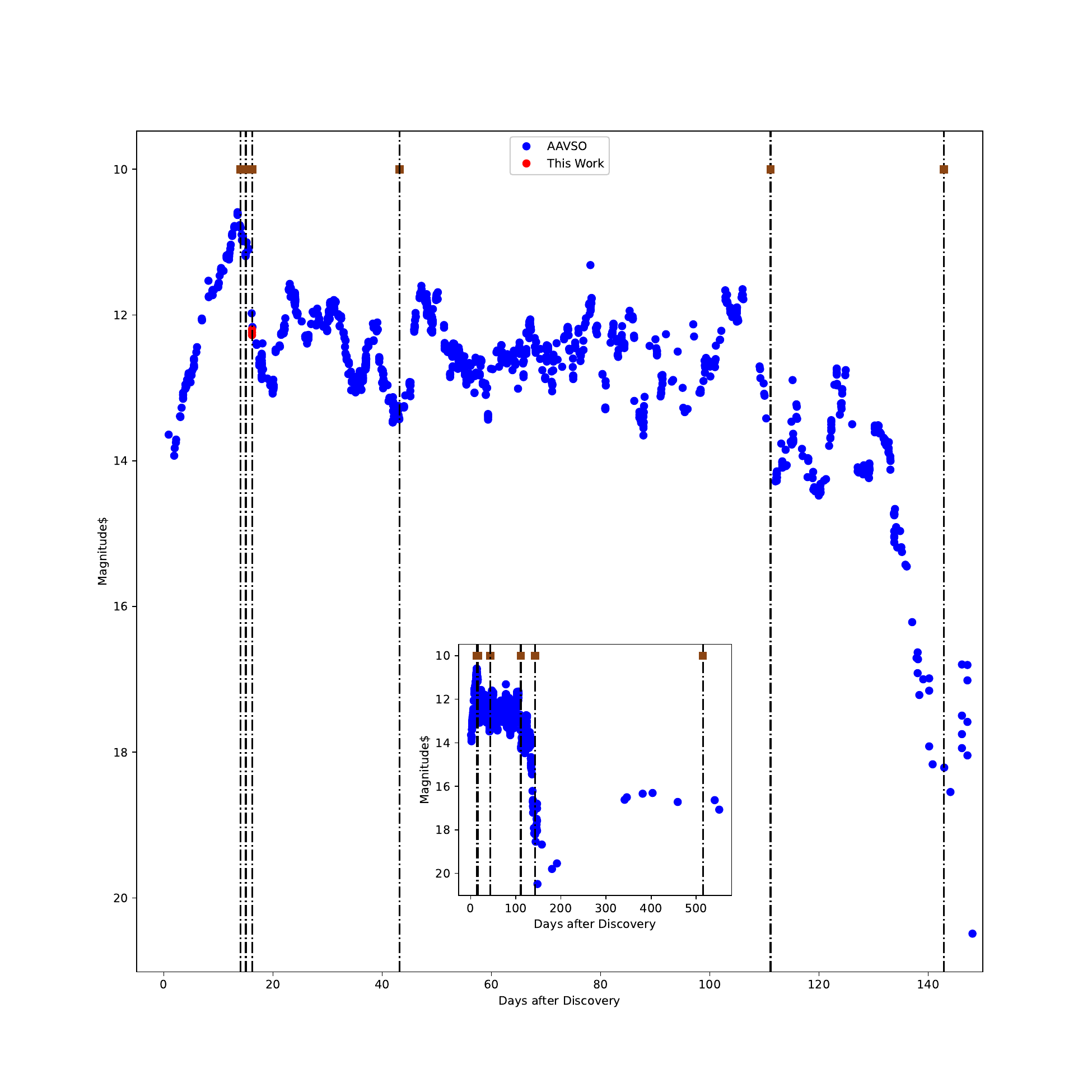}
\caption{V1391 Cas $V$-band observations (blue circles) from the AAVSO database. Only the 3 photometric data (red circles) belong to the RTT150. The times of the spectroscopic observations are labeled over the light curve with brown-filled squares and vertical dash-dotted lines. There are shown the phases of 17 spectra taken over nearly 500 days on the light curve. Accordingly, Days 14, 15, and 16 were closest to the maximum light, while \textbf{Day 43} was in the post-outburst multiple maxima. \textbf{Day 111} is in the decline just after the multiple maxima, and \textbf{Day 143} is in the decline before the transitional (nebular) phase. Finally, \textbf{Day 515} corresponded to the afterglow minimum caused by dust from the transitional phase. The larger plot shows the light curve from the discovery until the dust formation epoch while the smaller plot shows the entire light curve.}
\label{fig:aavso_v}

\end{figure}

\begin{figure}[ht]
\includegraphics[width=\textwidth]{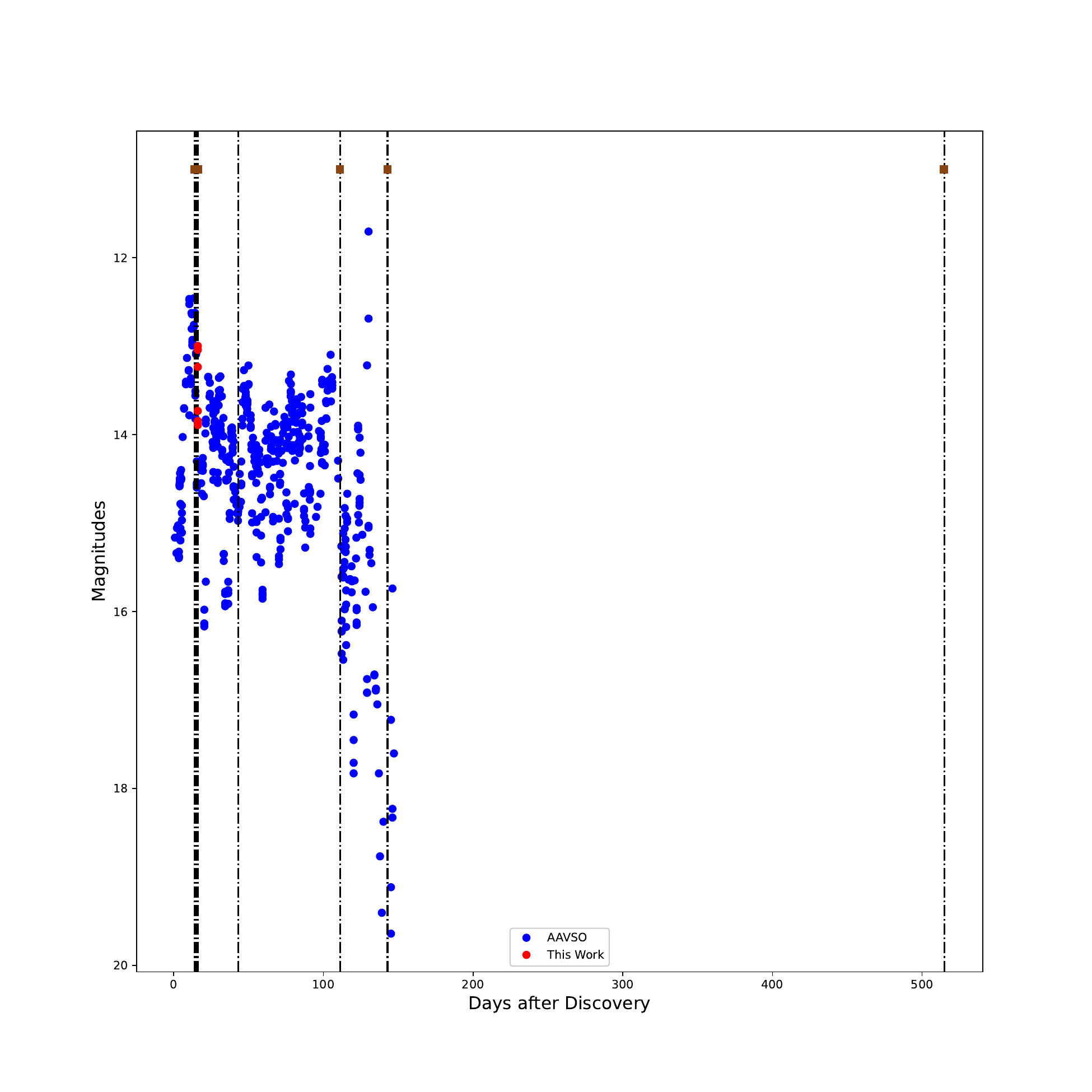}
\caption{Same as Figure \ref{fig:aavso_v} but for the $B$-band. Only the 6 photometric data (red circles) belong to the RTT150. Here, there are shown the phases of 17 spectra taken over nearly 500 days on the light curve. Accordingly, \textbf{Days 14}, \textbf{15}, and \textbf{16} were closest to maximum light, while \textbf{Days 43} and \textbf{111} were in the post-outburst multiple maxima. \textbf{Day 143} is in the decline after multiple maxima. It seems data missing in the AAVSO light curve from \textbf{Day 143} to \textbf{Day 515} due to dust in the transitional phase.}
\label{fig:aavso_b}

\end{figure}

Our first three nights of spectroscopic and photometric observations coincide with the visual maximum and the rapid magnitude decrease following it while our last spectroscopic observations are obtained after the nova's magnitude has risen after long dip due to dust formation (see Figs. \ref{fig:aavso_v} \& \ref{fig:aavso_b}). The photometric data with the blue circles in Figures \ref{fig:aavso_v} and \ref{fig:aavso_b} were taken from the AAVSO database \footnote{\textbf{(\url{https://www.aavso.org})}}. In addition, the 9 data points of 3 $V$ and 6 $B$-bands indicated by the red circles were taken by the RTT150 telescope with 10s exposure close to the maximum light. Therefore it would be interesting to see the spectral behavior of the nova at these epochs. The same Andor iKon-L CCD camera was used for the photometric observations with a field of view of $11.1^{'}$ x $11.1^{'}$. For convenience, the times of the spectroscopic observations are labeled in Figures \ref{fig:aavso_v} and \ref{fig:aavso_b} over the AAVSO light curves.

\begin{figure}[ht]

\includegraphics[width=\textwidth]{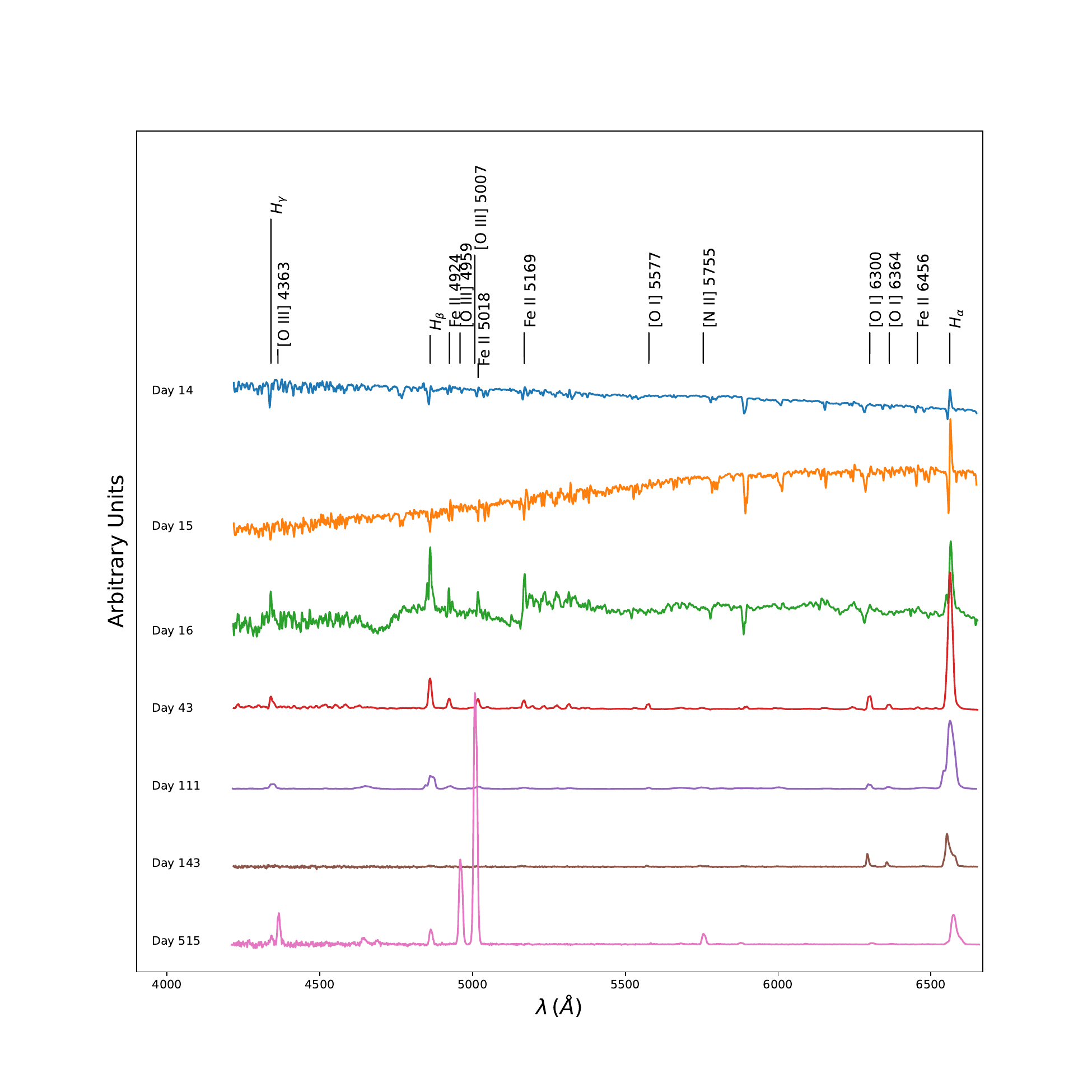}
\caption{V1391 Cas observations in the blue part 4220-6650 $\si{\angstrom}$ with 1331 resolution (grism 7). In this region and at this resolution, the evolution of the spectra (trend of the continuum, change of spectral lines) is seen for about 500 days after the discovery of the nova (from \textbf{Day 14} to \textbf{Day 515}).}
\label{fig:g7}

\end{figure}

\begin{figure}[ht]
\includegraphics[width=\textwidth]{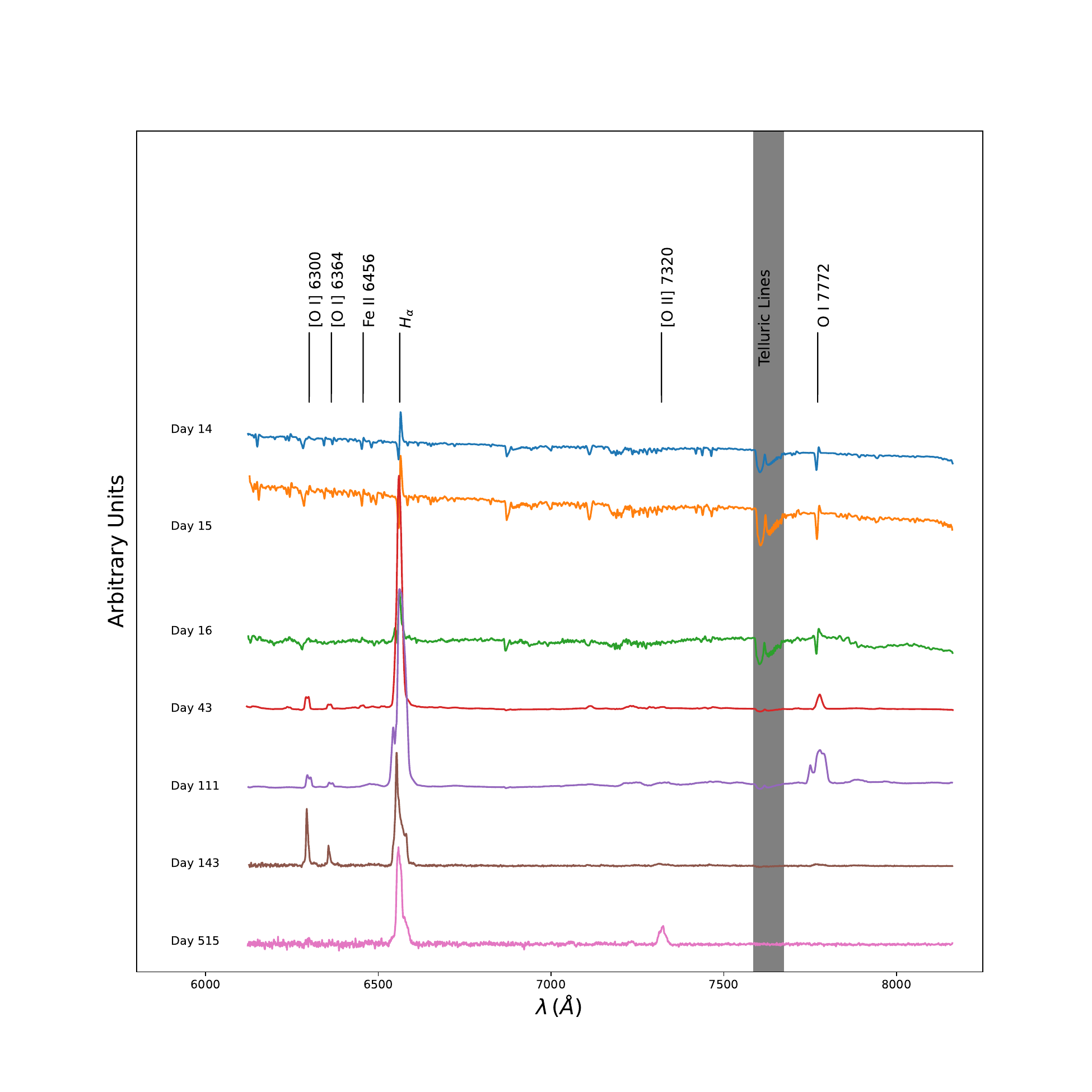}
\caption{V1391 Cas observations in the red part 6190-8190 $\si{\angstrom}$ with 2189 resolution (grism 8). In this region and at this resolution, the evolution of the spectra (trend of the continuum, change of spectral lines) is seen for about 500 days after the discovery of the nova (from \textbf{Day 14} to \textbf{Day 515}).}
\label{fig:g8}

\end{figure}

\begin{figure}[ht]
\includegraphics[width=\textwidth]{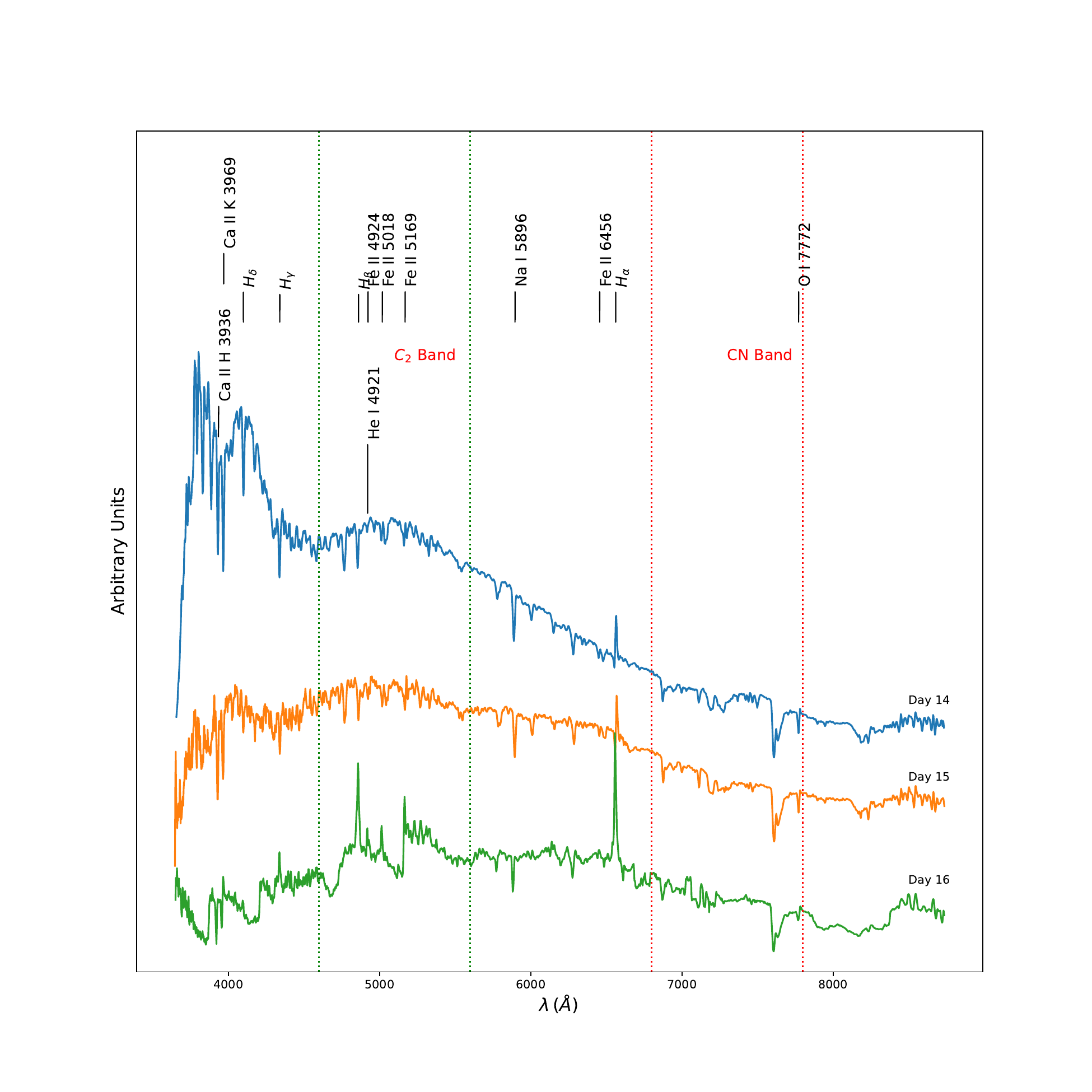}
\caption{V1391 Cas observations in the blue-red part 3650-8740 $\si{\angstrom}$ with 749 resolution (broad-band grism 15). In this region and at this resolution, the evolution of the spectra (trend of the continuum, change of spectral lines) is seen for about 3 days after the discovery of the nova (from \textbf{Day 14} to \textbf{Day 16}).}
\label{fig:g15}

\end{figure}

The spectra taken using the three grisms are plotted in Figs.  \ref{fig:g7}, \ref{fig:g8}, and \ref{fig:g15}. The common aspects of the spectra given in the figures are as follows: 1) The spectra are indicated with different colors. 2) Some spectral lines are also marked. 3) The days of the spectra taken after the outburst are shown on the V and B-bands light curves with dash-dotted lines (see Figures \ref{fig:aavso_v} and \ref{fig:aavso_b}).


%\subsection{Data reduction}

The basic data reduction was made using the Image Reduction and Analysis Facility (IRAF) version 2.16 \citet{1993ASPC...52..173T} and the task \textit{splot} was used for measuring the spectral lines. The multiple peaked lines were deblended using the deblending method of splot where gaussian profiles were fitted for each peak. Many frames were heavily affected by cosmic rays so the \textit{LACosmic} routine of \citet{2001PASP..113.1420V} was used for cosmic ray removal making use of its Astro-SCRAPPY version \citep{curtis_mccully_2018_1482019}. Spectrophotometric standard stars from \citet{1990AJ.....99.1621O} were used for the flux calibration. These stars are listed in the sixth column of Table ~\ref{tab:log}. The standard method for flux calibration in IRAF was used. All the spectra were corrected for interstellar extinction adopting a value of $E(B-V) = 1.39$ calculated by \citet{2020ATel13905....1M} adopting an $R_V$ value of 3.1. This value is consistent with the extinction values obtained from the interstellar dust maps of \citet{1998ApJ...500..525S} and \citet{2011ApJ...737..103S}. This value is also consistent with the value calculated for the extinction of supernovae by \citet{2016ApJ...826...66F}. The dust\_extinction tool included in {\sc Astropy} \citep{astropy:2013,astropy:2018} using the interstellar extinction curves of \citet{2019ApJ...886..108F} was used to correct for interstellar extinction. The methods described by \citet{1992PASP..104.1104L} were used to calculate the errors in the line fluxes and velocities, while  \citet{1988IAUS..132..345C} method was used to calculate the uncertainty in equivalent widths determinations. The {\sc Astropy} affiliated package {\it photutils} was used for calculating the magnitude in the photometric observations \citep{larry_bradley_2020_4044744}.
For the 2021 observations, we used the {\sc Astropy} affiliated package {\sc ccdproc} \citep{matt_craig_2017_1069648} for image reduction and a special python routine developed by TUG staff \citep{2020AstL...46....1K} was used for spectrum extraction and wavelength calibration.

The nova was very faint on \textbf{Day 143} and only a few emission lines were visible. This is consistent with the dust formation reported by \citet{2020ATel14272....1B}. On Day 162, the nova was detectable using direct imaging but it was too faint to produce a good spectrum using RTT150. We were able to observe the nova again on \textbf{Day 515} when the dust had dissipated. It should be noted that there is no information in the literature on nova evolution for \textbf{days 143} and \textbf{515} in Section 1.1.

\begin{figure}
\centering
    \begin{subfigure}[t]{0.45\textwidth}
        \centering
        \includegraphics[width=0.86\linewidth]{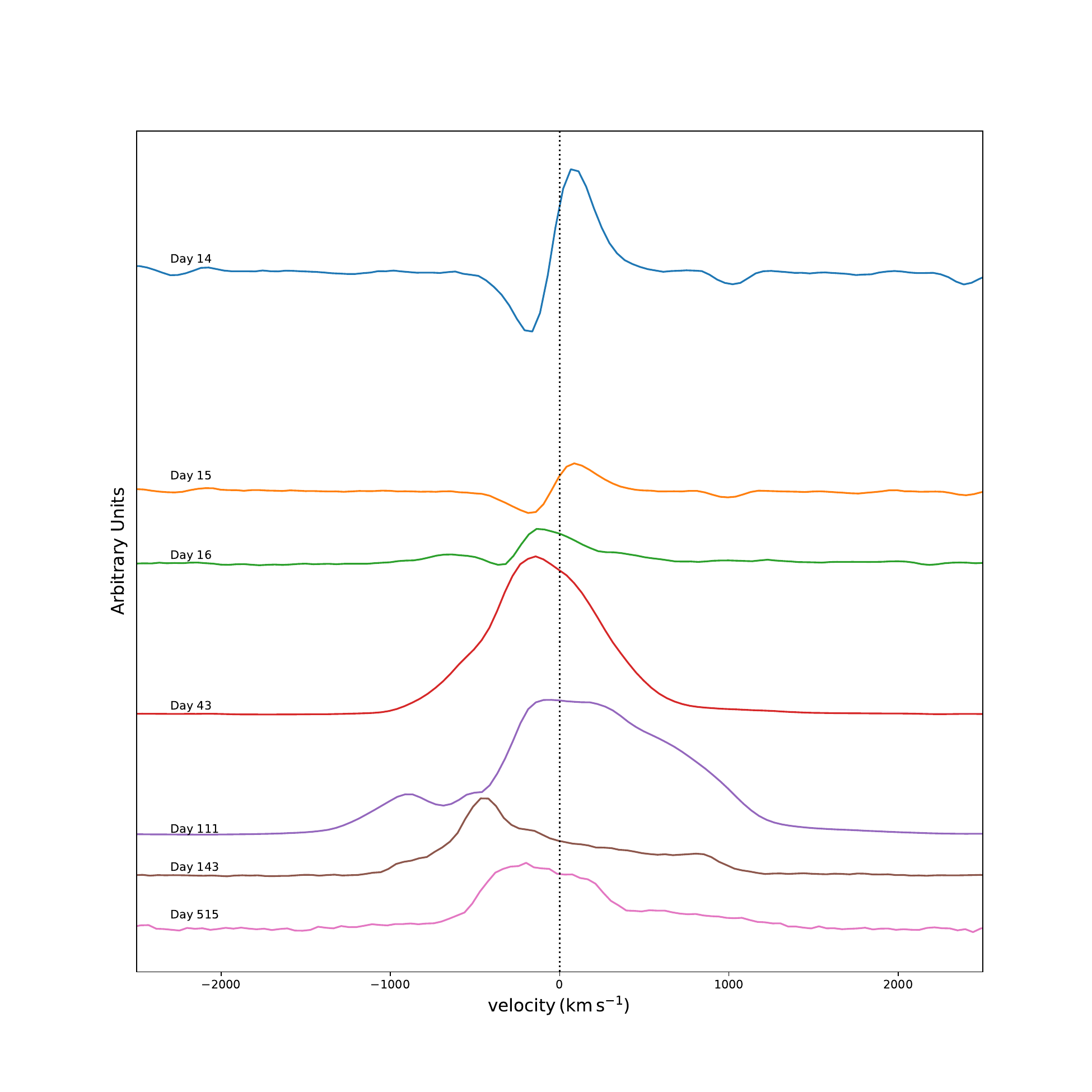} 
        \caption{H$\alpha$.} \label{fig:halpha_g8}

    \end{subfigure}
    \hfill
    \begin{subfigure}[t]{0.45\textwidth}
        \centering
        \includegraphics[width=0.86\linewidth]{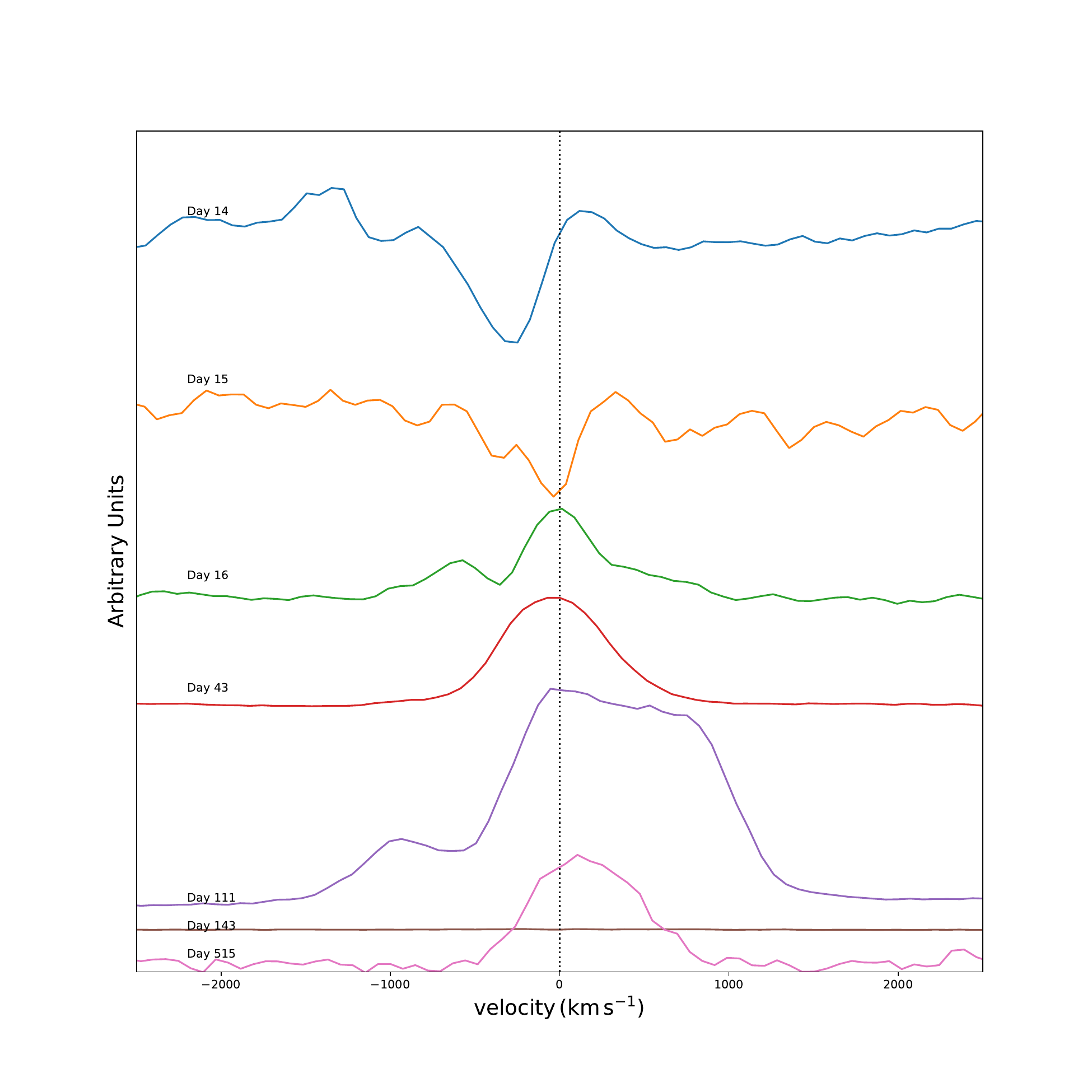} 
        \caption{H$\beta$.} \label{fig:hbeta_g7}
    \end{subfigure}
    \vspace{1cm}
    \begin{subfigure}[t]{0.45\textwidth}
    \centering
        \includegraphics[width=0.86\linewidth]{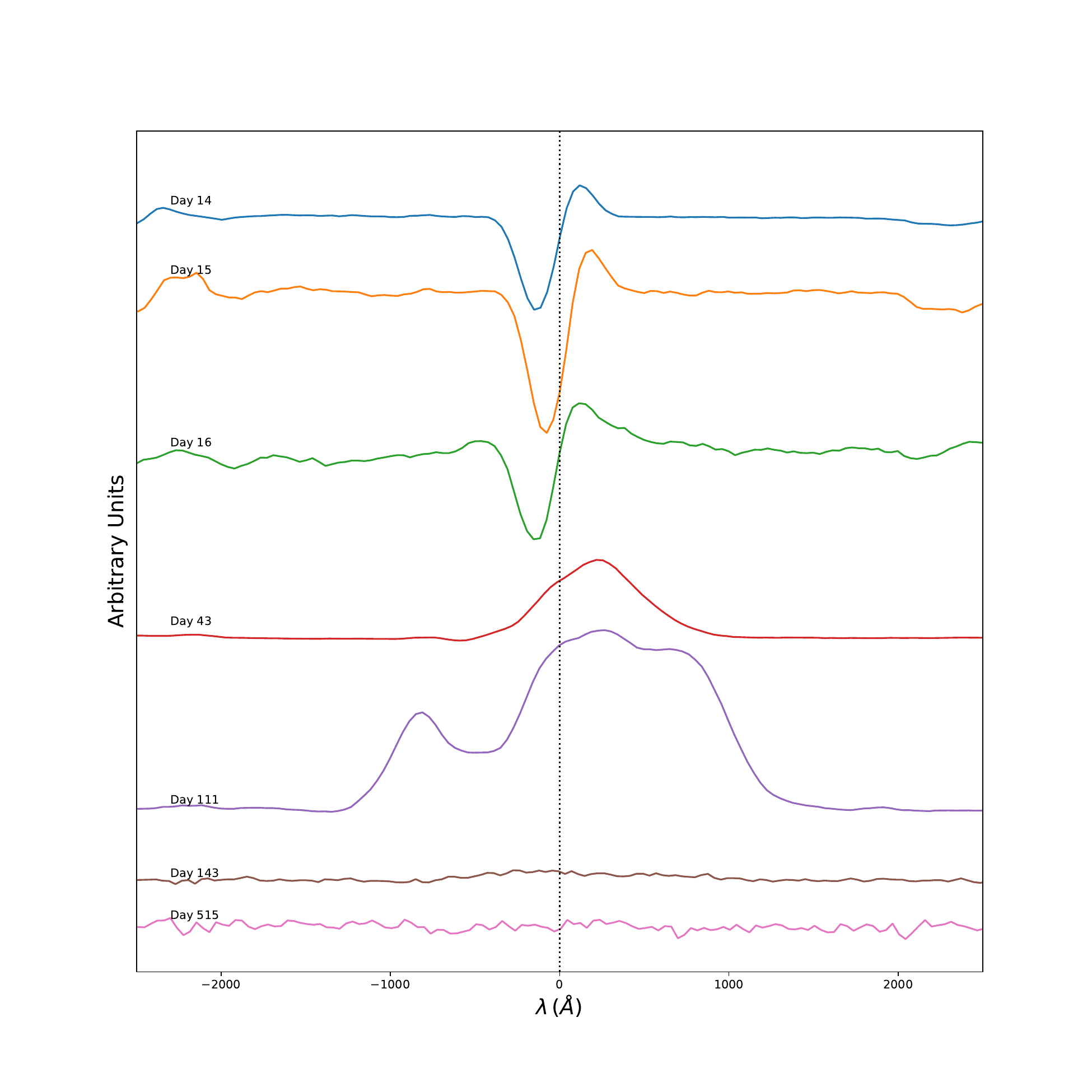} 
        \caption{ O I 7772 $\si{\angstrom}$.} \label{fig:oi_g8}
    \end{subfigure}
    \hfill
        \begin{subfigure}[t]{0.45\textwidth}
        \centering
        \includegraphics[width=0.86\linewidth]{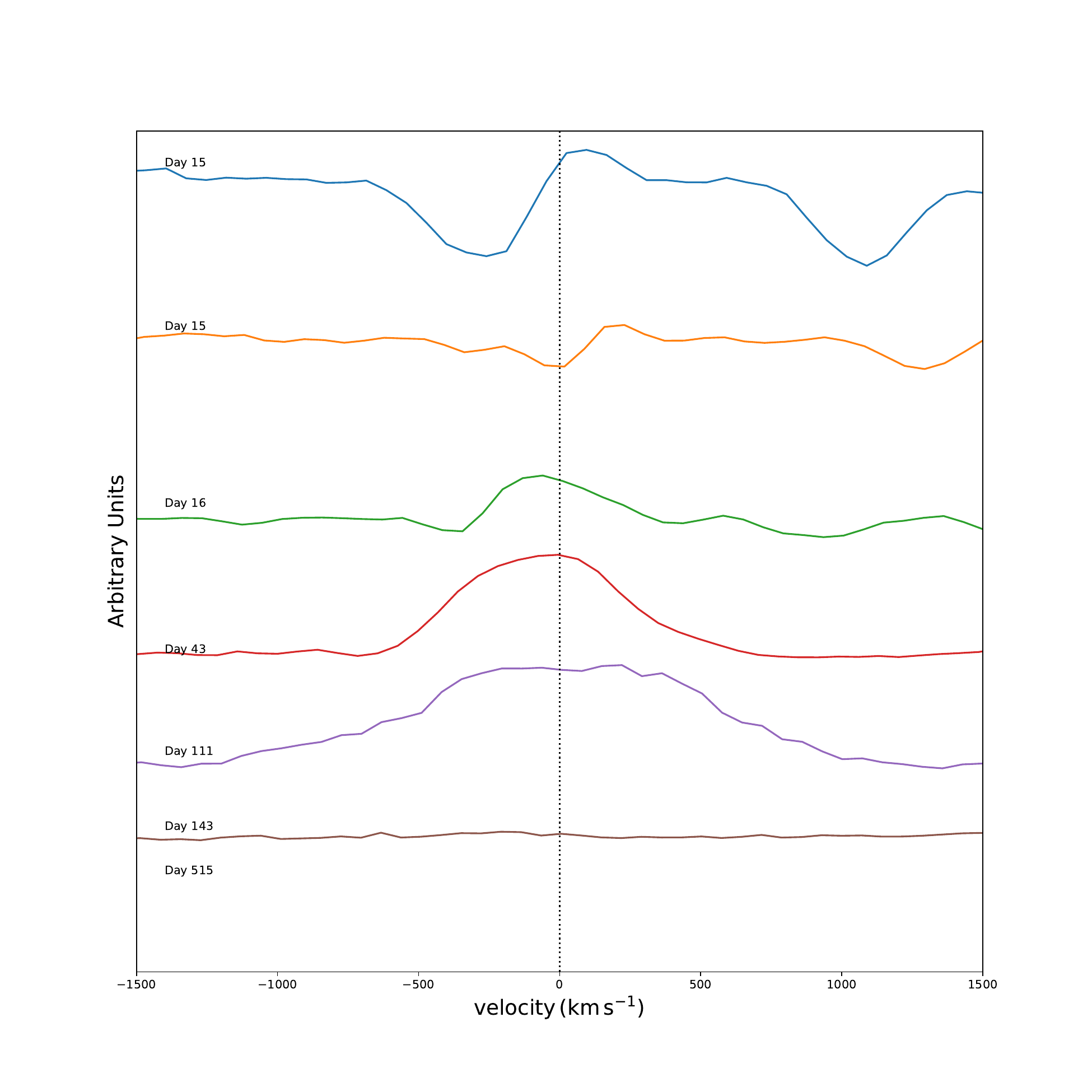} 
        \caption{ Fe II 5018 $\si{\angstrom}$. } \label{fig:fe_ii_g7}
    \end{subfigure}
    \begin{subfigure}[t]{0.45\textwidth}
        \centering
        \includegraphics[width=0.86\linewidth]{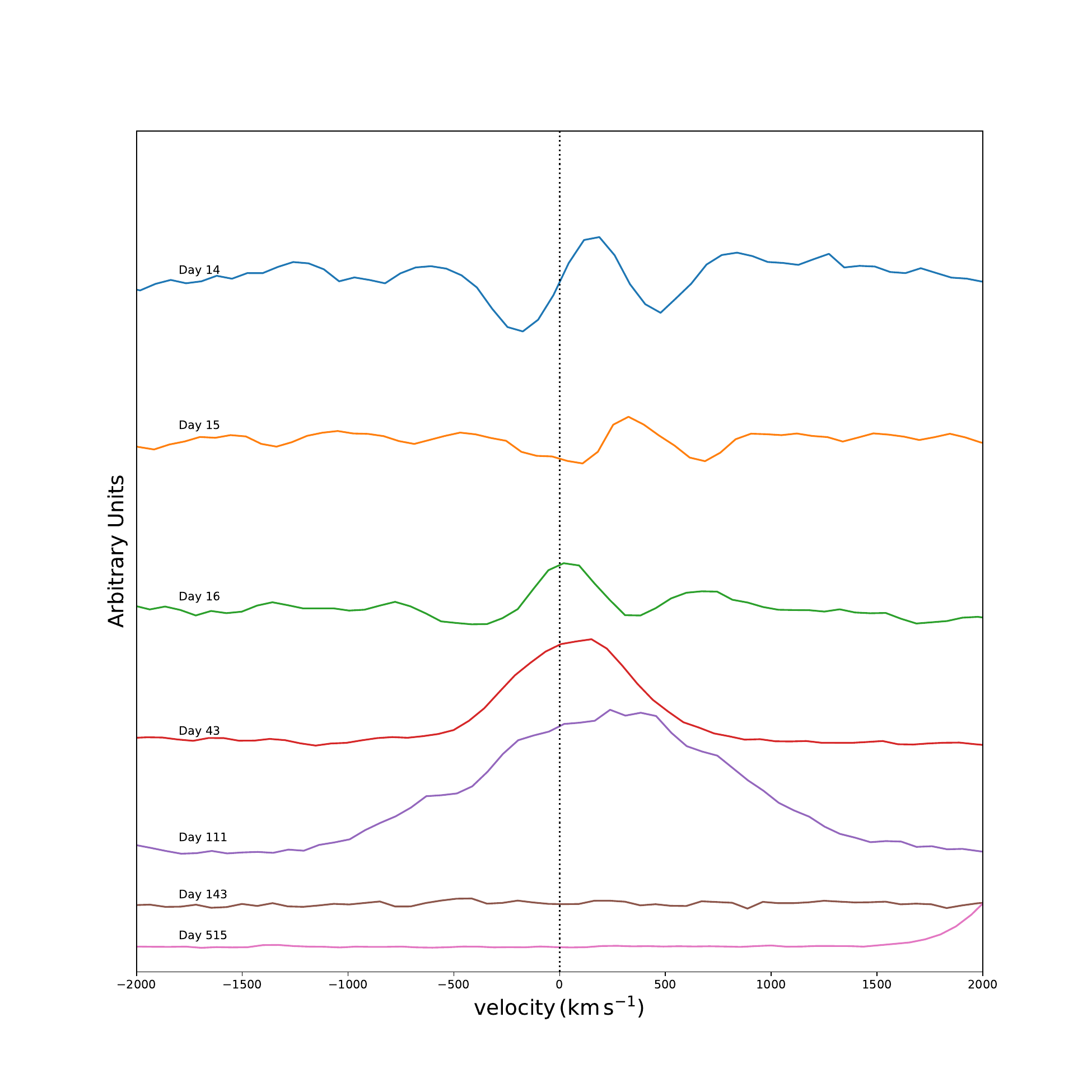} 
        \caption{He I 4921 $\si{\angstrom}$.} \label{fig:he_i_g7}
    \end{subfigure}
    \caption{Spectral evolution of some prominent features.}
\end{figure}

\begin{figure}[ht]
\includegraphics[width=0.9\textwidth]{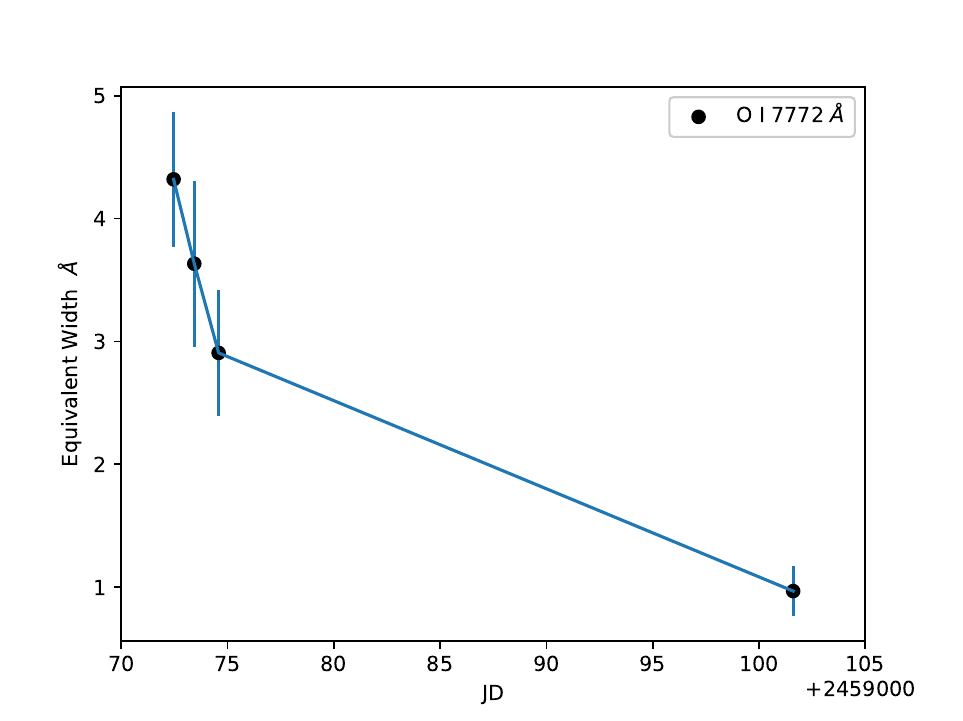}
\caption{Evolution of the equivalent width of absorption component of O I 7772$\si{\angstrom}$ line. }
\label{fig:eq_ab_oi}

\end{figure}

\begin{figure}[ht]
\includegraphics[width=0.9\textwidth]{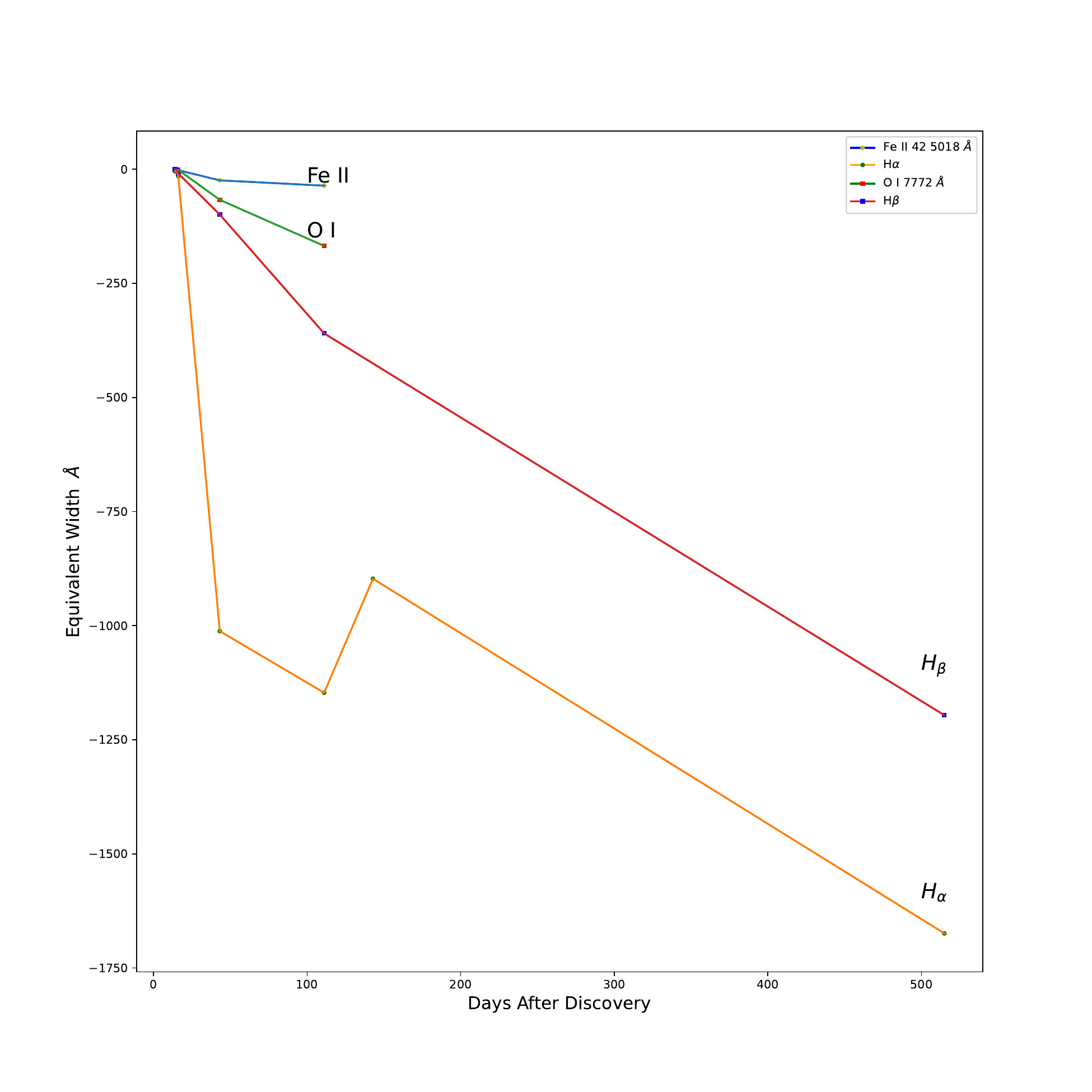}
\caption{Evolution of the equivalent width of emission components of some prominent lines. }
\label{fig:eq_em}

\end{figure}

%\section{Results}
%\subsection{Spectroscopy}
We made use of the NIST database of \citet{NIST_ASD} to determine the rest wavelengths of spectral lines. The spectral evolution of some prominent lines shows that the H$\alpha$ line started with a clear P-Cygni profile with the absorption feature at about -200 $\mathrm{km\,s^{-1}}$ on \textbf{Day 14} (see Table \ref{tab:absorption}, Day 14). The observations taken in the following days showed that the line got weaker in both absorption and emission (see Table \ref{tab:absorption}, \textbf{Day 15} and \textbf{16}, and also Figure \ref{fig:eq_em}). These spectra showed also that the line was weak but blue-shifted (see Figure \ref{fig:halpha_g8}). In the observations taken on \textbf{Day 43}, the absorption component almost disappeared and the line was completely in emission. By \textbf{Day 111} the line reached its maximum flux in our observations and it showed a double-peaked profile with the red peak having the higher flux. On \textbf{Day 143} spectra, the line showed multiple peaks while the blue peak had the highest flux (see Figs. \ref{fig:halpha_g8} and \ref{fig:eq_em}, Tables \ref{tab:absorption} and \ref{tab:emission}). The H$\beta$ line showed an absorption feature in \textbf{Day 14} observations. Then the emission component grew stronger and the absorption component got blue-shifted in the \textbf{Day 15} observations. By the time of our observations on \textbf{Day 43}, the line transformed completely to emission. It reached its maximum flux in \textbf{Day 111} spectra and for \textbf{Day 143} spectrum, it was undetected. It reappeared in emission in \textbf{Day 515} observations (see Table \ref{tab:emission} and Figs. \ref{fig:hbeta_g7}, and \ref{fig:eq_em}). The evolution of the O I 7772 $\si{\angstrom}$ line showed very similar behavior to the evolution of the H$\alpha$ line where the line shows a P-Cygni profile with the absorption component blue-shifted at a velocity of about -130 $\mathrm{km\,s^{-1}}$. In a similar behavior to that of H$\alpha$, the line got weaker in both absorption and emission in \textbf{Day 15} observations. In \textbf{Day 16} spectra the line was blue-shifted while remaining weak. In \textbf{Day 43} spectra, the absorption component nearly disappeared and the line was in emission (see Figs. \ref{fig:oi_g8}, \ref{fig:eq_ab_oi}, and \ref{fig:eq_em}). This similarity was also noted in the behavior of both H$\alpha$ and O I 7772 lines in the spectra of the classical nova V659 Sct \citep{2020AN....341..781J}. However, the flux of the emission line of H$\alpha$ was much stronger in the spectrum taken on \textbf{Day 43}. 

\begin{table}
\caption[]{The evolution of the equivalent widths of the absorption component of some lines. The three dots (...) indicate that the line was undetected.}    
\label{tab:absorption}
%\begin{center}
\begin{tabular}{lcccc}
\hline \hline
Element    &            Day 14 & Day 15 & Day16 &  Day 43 \\
\hline
H$\alpha$ &  $2.2\pm 0.9$                 & $3.1\pm 0.3 $       &       &         \\
H$\beta$  &  $2.1 \pm 0.2$                 &   $1\pm 1  $      &       &         \\
%H$\gamma$  &  $39 \pm 5$               &         &       &         \\
O I 7772   &  $3.7\pm 0.8$                 &  $3.6\pm 0.7 $      &  $2.9\pm 0.5$     &   $0.6\pm 0.3 $      \\
Fe II 4924 &  $0.5\pm 0.2$                 & $2.0\pm 0.3$       &  $1.0 \pm 0.2$     &          \\
      5169 &      $1.1 \pm 0.2 $              &  $1.2 \pm  0.7 $      &        &          \\
      5232 &      $0.3 \pm 0.2 $              &         &        &          \\
      6456 &  $1.1\pm 0.2  $                 &  $1.1\pm 0.2  $      &       &          \\
\hline
\end{tabular}
%\end{center}
\end{table}

%Table 1 needs to be reviewed (especially snr)
%I have to review the plots (except those for the velocity component)

\begin{landscape}

\begin{table}[ht]
\centering
\caption{The evolution of the integrated flux of the emission component of some lines (in $\mathrm{10^{-11}\,erg \, cm^{-1} \,s^{-1}}$) .} 
\label{tab:emission}
%\begin{center}
\begin{footnotesize}

\begin{longtable*}{rccccccc}
\hline \hline
Element       &     Day 14 & Day 15 & Day16 &  Day 43& Day 111 &Day 143 & Day 515 \\
\hline
H $\alpha$ &  $4.4\pm 0.5$          &  $0.60\pm 0.09 $      & $1.24 \pm 0.06 $   & $107\pm 2 $    & $8.0 \pm    0.3$*    & $0.16  \pm 0.03  $** & $0.60  \pm 0.03  $  \\
 &            &                    &                        &                    & $141 \pm  2  $     & $0.72  \pm 0.02  $ &  \\
 &            &                    &                        &                    &                    & $0.218  \pm 0.006  $  & \\
H $\beta$ &   $0.14\pm 0.02 $ &      $0.15 \pm 0.02$ & $2.1\pm 0.3  $       & $10 \pm 1  $   & $4.5 \pm 0.3  $ **   & &$0.48 \pm 0.06 $     \\
          &                   &                      & $0.62 \pm 0.01  $    &                &      $19.7 \pm 0.3$  &    \\
          &                   &                      &                      &                &      $17.5 \pm 0.3$  &    \\
H$\gamma$  &            &        & $0.9 \pm 0.1 $    &  $2.5 \pm 0.3$       & $14.5 \pm 0.6$     &    & $0.23\pm 0.07$\\
           &            &        &                   &  $0.97 \pm 0.03$     &                    &    \\
           &            &        &                   &  $1.29 \pm 0.04$     &                    &    \\
O I 7772  &  $0.71\pm 0.08$          &   $0.100 \pm 0.009 $     & $0.089 \pm 0.0007$   & $6.2 \pm 0.1 $     &  $18.40\pm 0.02  $ *   &   & \\
          &                          &                          & $0.081 \pm0.0001$    &                    & $9.6 \pm 0.1 $     &    & \\
          &                          &                          &                      &                    &                     &    & \\
Fe II 4924  &            &        & $0.5 \pm 0.1 $   & $3.3\pm 0.2$    &   $9.1 \pm 0.2 $   &    &    \\
        & $0.32 \pm 0.02$           &        & $0.67\pm 0.06$   & $2.8\pm 0.2$    &  $5.5 \pm 0.1 $    &    &  \\
      5169  &  $0.56 \pm 0.03 $          & $0.45 \pm 0.03$       & $1.6\pm  0.1$   & $2.6\pm  0.2$    &  $3.1 \pm 0.1$    &    & \\
      6456  &     $0.34 \pm 0.03 $ &$0.036 \pm 0.0003 $      &        & $0.38\pm 0.01 $*   &   $3.27\pm 0.01$  &      &    \\
            &                      &                         &        & $0.47\pm 0.01 $    &                   &      &    \\
      5235  &   $0.80 \pm 0.01 $         & $0.105 \pm 0.003 $       &  $0.94 \pm 0.04 $  &   $0.423\pm 0.007  $  &      &   & \\
      5276  &                  &        &    &                     &  $1.26 \pm 0.02 $    &   & \\
{[}N II{]} 5755   &            &        &    &                    &                     &    &$ 0.41 \pm 0.03$\\
{[}O III{]} 4363   &            &        &    &                    &                     &    &$ 0.8 \pm 0.1$ \\
            4959   &            &        &    &                    &                     &    &$ 3.1 \pm 0.3$\\
            5007   &            &        &    &                    &                     &    &$ 9 \pm 1$\\
{[}O I{]} 5577    &            &        &    & $1.7 \pm 0.1 $ *  & $1.8 \pm 0.2 $     & $0.10 \pm 0.02 $   &\\
          6300    &            &        &    & $2.3 \pm 0.1 $  *  & $4.9 \pm 0.3 $ *    & $0.5 \pm 0.1$   & $0.068 \pm 0.004$\\
                  &            &        &    &  $2.53 \pm 0.08  $   &  $4.8 \pm 0.3 $    &    &\\
          6364    &            &        &    & $0.78 \pm 0.03$ *   & $1.84 \pm 0.08 $  *   & $ 0.15 \pm 0.03$   &$0.036 \pm 0.004$\\
                  &            &        &    &  $0.80 \pm 0.03 $  &  $1.58\pm 0.06 $    &    &\\
{[}O II{]} 7320   &            &        &    &                    &                     &    &$ 0.123 \pm 0.007$\\
\hline

%\caption*{* Double Peaked, ** Multiple Peaked}
$^*$ Double Peaked\\
$^{**}$ Multiple Peaked
\end{longtable*}
\end{footnotesize}
%\end{center}
\end{table} 

\end{landscape}

On \textbf{Day 16} low-resolution spectra, CN and $\mathbf{C_2}$ bands are clearly visible. This was also noted by \citet{2021ApJ...907...70F} for this nova.

The He I 4921 $\si{\angstrom}$ was one of the prominent lines observed in our spectra. It was not observable in \textbf{Day 14} spectra. It started to show a P-Cygni profile on \textbf{Day 15}. By \textbf{Day 43} the absorption component disappeared and the line was completely in emission (similar to the behavior of the profiles of H$\alpha$ and H$\beta$ lines). It reached the maximum flux in \textbf{Day 111} observations and by \textbf{Day 143} it was not observable (see Fig. \ref{fig:he_i_g7}).

\begin{figure}[ht]
\includegraphics[width=0.9\textwidth]{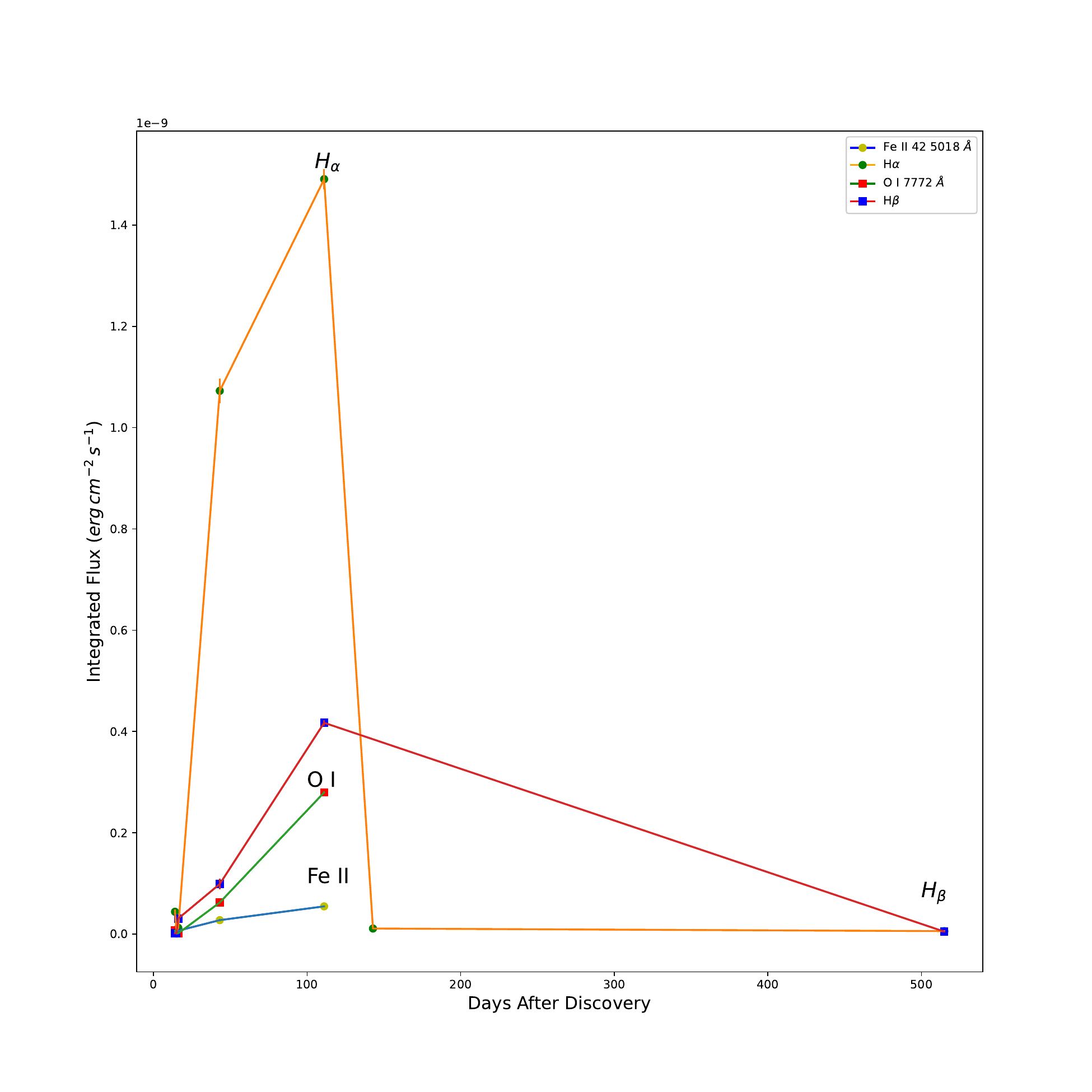}
\caption{Evolution of the flux of emission components of some prominent lines. }
\label{fig:flux_em}

\end{figure}

Numerous Fe II lines are seen in the spectra. The most prominent of these lines are the 42 multiplet lines at the wavelengths 4924 $\si{\angstrom}$,  $\si{\angstrom}$, and 5169 $\si{\angstrom}$. These three lines showed P-Cygni profiles for the first two nights (Days 14 and 15). Then they showed only emission components reaching maximum fluxes on \textbf{Day 111}. Similar behavior was shown by the 5235 $\si{\angstrom}$, 5276 $\si{\angstrom}$, and 5317 $\si{\angstrom}$ Fe II lines. The fluxes are presented in Tables \ref{tab:absorption} and \ref{tab:emission}. The 6456 $\si{\angstrom}$ Fe II line was seen in absorption on \textbf{Day 14}-\textbf{15} low-resolution spectra and it transitioned to P-Cygni profile for \textbf{Day 15} low-resolution spectrum. The evolution of the Fe II  $\si{\angstrom}$ line is presented as a sample of Fe II lines in Fig. \ref{fig:fe_ii_g7}. Fig. \ref{fig:flux_em} shows the evolution of the flux of some prominent emission lines.

On \textbf{Day 515} observations, some higher ionization lines such as [O III] 4363 $\si{\angstrom}$, [O III] 4959 $\si{\angstrom}$, [O III] 5007 $\si{\angstrom}$, [O II] 7320 $\si{\angstrom}$ and [N II] 5577 $\si{\angstrom}$ appeared in the spectra (see Table \ref{tab:emission}  for the fluxes of these lines).

\begin{figure}[ht]
%\centering
\includegraphics{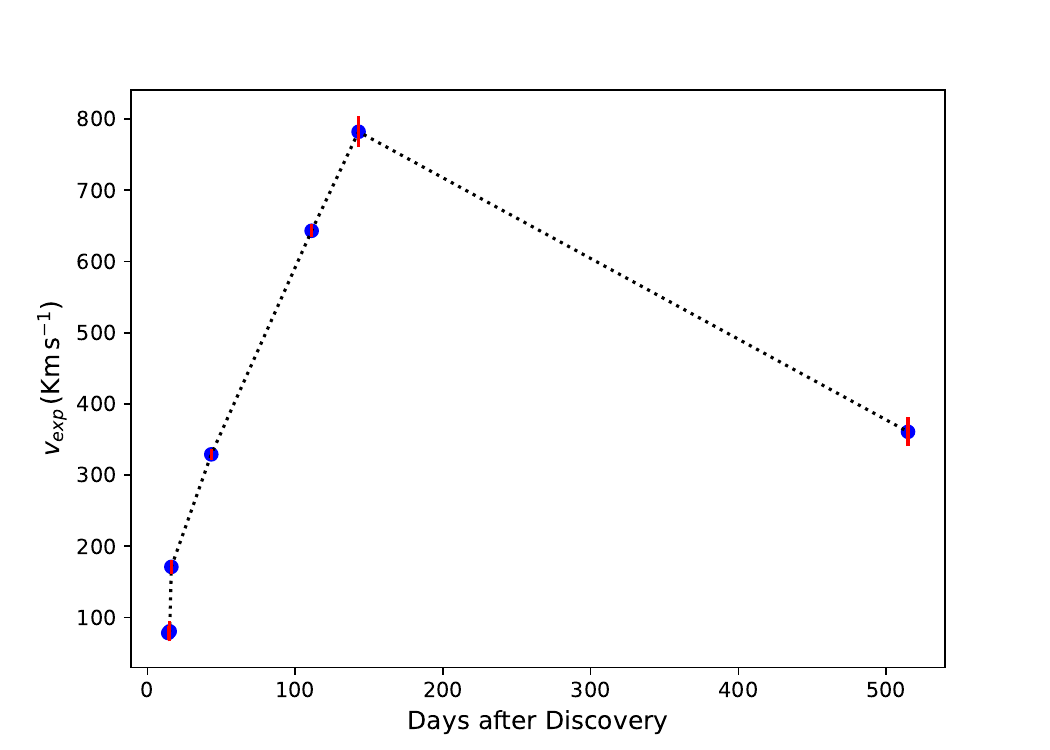}
\caption{The expansion velocity of V1391 Cas using the H$\alpha$ FWHM.}
\label{fig:vexp}
\end{figure}

%Only one component of the interstellar absorption Na I D doublet was observed during the Aug $10^{th}\,-\,12^{th} $ observations. The Ca H line was observed during the observations while the Ca K line was observed during the observations. The K I interstellar doublet was observable only during the Aug $10^{th} $ observations (see Table \ref{tab:interstellar}).
%\clearpage
The expansion velocity of the envelope was calculated from the FWHM of the emission component of the H$\alpha$ line (corrected for instrumental resolution) using the same method used by \citet{2010PASP..122..898M}. The expansion velocity has risen from $\sim$ 80 $\mathrm{km\,s^{-1}}$ on \textbf{Day 13} to $\sim$ 780 $\mathrm{km\,s^{-1}}$ on \textbf{Day 143}. It has decreased again to $\sim$ 360 $\mathrm{km\,s^{-1}}$ on \textbf{Day 515}. A plot of the evolution of the expansion velocity is shown in Fig. \ref{fig:vexp}.

%\begin{table}
%\label{tab:FeII_fluxes_em}
%\caption{Flux of the emission component of  Fe VI line}
%\caption{Evolution of the emission component of Fe II line}
%\begin{longtable}{ccc}
%\hline \hline
%Fe VI 5677 &169.5321 & $6.47 \pm 0.03 \times 10 ^{-11}$ \\
%\hline
%\end{longtable}
%\end{table}

%\addtocounter{table}{-1}

%\begin{longtable}{rccc}
%\caption{Equivalent widths of some interstellar absorption lines.}
%\label{tab:interstellar}

%\hline \hline 
%Element & JD (2459000+)  & EW ($\si{\angstrom}$)     \\
%\hline
%Ca H 3934 &072.40733  & $4.6 \pm 0.6 $    \\
%&073.30314       & $5 \pm 1$\\
%&074.57196  &    &  \\

%Ca K 3969 &072.40733  & $5 \pm 2$   \\
%&073.30314  &    &  \\
%&074.57196  &    &  \\

%Na I 5892& 072.40733  &$4.1 \pm 0.4$    \\
%&073.30314  & $4.8 \pm 0.3$     \\
%&074.57196  &  $3.9 \pm 0.5$    \\
%\hline
%\end{longtable}

%K I 7665& 072.4635 &  $1.0 \pm 0.2$ &  \\

 %   7699 &072.4635 &  $0.4 \pm 0.1$ &  \\

In the spectrum of \textbf{Day 43}, the forbidden lines [O I] 5577 $\si{\angstrom}$ along with the [O I] $\lambda\lambda$ 6300 $\si{\angstrom}$, 6364 $\si{\angstrom}$  doublet started to show in the spectra. These lines were observable until \textbf{Day 143}. The Integrated fluxes are included in Table \ref{tab:emission}. %while the equivalent widths are included in Table \ref{tab:oi_f}.

\begin{table}
\caption{Evolution of the equivalent width of {[}O I{]} forbidden lines.}

\label{tab:oi_f}

\begin{longtable}{lcl}
\hline \hline

Day & Element\  & EW ($\si{\angstrom}$)\\
\hline
43&{[}O I{]} 5577 &$20.6\pm 0.6$\\
111&  &$14.5 \pm 0.7$\\
143 &  &$86 \pm 1$\\
\hline
43*& {[}O I{]} 6300 &$31.0\pm 0.2$\\
43&  &$32.3\pm 0.1$\\
111* &   &$40.14 \pm 0.3$ \\
111&  &$38.1 \pm 0.3$ \\
143&   & $293 \pm 1$ \\
515 &  & $97.6 \pm 0.6$  \\
\hline
43*& {[}O I{]} 6364 &$8.9\pm 0.2$\\
43&  &$8.9\pm 0.1$\\
111* &   &$12.3 \pm  0.2$ \\
111 &  &$10.2 \pm 0.2$ \\
143 &   & $53 \pm 1$\\
515 &  &$51 \pm 1$\\
\hline
\caption*{* double-peaked.}
$^*$ Double Peaked
\end{longtable}
\end{table}

\section{The Mass of The Ejected Envelope}
\subsection{The mass of neutral oxygen in the ejecta}

Using the method described by \citet{1994ApJ...426..279W}, the ratio of the fluxes of the  6300 $\si{\angstrom}$ to 6364 $\si{\angstrom}$ [O I] lines (${F_{\lambda 6300}}/{F_{\lambda 6364}}$, see Table \ref{tab:physical_res}) was found to be almost 3$\colon$1 throughout the first three observations where this doublet was observable. This is identical to the theoretical value calculated from the ratio of the transition probabilities of this doublet \citep{1994ApJ...426..279W}. This result is in contrast to most of the novae which show values less than 3 \citep{1994ApJ...426..279W}. See Table \ref{tab:oi_f} for the equivalent widths of the [O I] lines.

%\clearpage
The ratio of the fluxes of this [O I] doublet (${F_{\lambda 6300}}/{F_{\lambda 6364}}$) was used to calculate the optical depth of the ejecta at $\lambda 6300$ using the equation

\begin{equation}
    \frac{F_{\lambda 6300}}{F_{\lambda 6364}} = \frac{1-e^{-\tau}}{1-e^{-\tau/3}}
\end{equation}

The calculated optical depth along with the ratio of the fluxes of the nebular line $\lambda 6300$ and the auroral $\lambda 5577$ line (equation 1) was used to calculate the electron temperature $T_e$ in the region where the [O I] lines are formed, where 

\begin{equation}
    T_e = \frac{11200}{log[43\tau\mathbin{/}(1-e^{-\tau})\times F_{\lambda 6300}\mathbin{/}F_{\lambda 5577}]}
\end{equation}

The electron temperature ($T_e$) with the flux of the auroral $\lambda 5577$ line (${F_{\lambda 5577}}$) was used to obtain the mass of neutral oxygen in the ejecta using the equation 

\begin{equation}
  M_{OI} = 152d^2\,e^\frac{22850}{T_e}\frac{\tau}{1-e^{-\tau}}\times F_{\lambda 6300}
    \end{equation}
where d is the distance in kiloparsecs and the flux (${F_{\lambda 6300}}$) is corrected for interstellar reddening (see \citealp{1994ApJ...426..279W} for equations).

\citet{2019A&A...622A.186S} concluded that the average absolute magnitude for classical novae 15 days after the maximum to be

\begin{equation}
    M_{V,15} = -5.71\pm 0.40
\end{equation}
the $V$ magnitude of V1391 Cas 15 days after the maximum was 12.16 which yields a distance of $\sim$5.2 kpc. Using this value we obtain the neutral oxygen masses tabulated in Table \ref{tab:physical_res}.

All the values of $F_{\lambda 6300}/F_{\lambda 6364}$ are almost equal to 3 therefore we adopt a value of 2.84 for the calculations of $\tau_{\lambda6300}$, $T_e$  and $M_{OI} $. Our calculations (see Table \ref{tab:physical_res}) give an average electron temperature $\mathbf{T_e = 4890\,K}$ and an average neutral oxygen mass $\mathbf{M_{OI} = 2.54  \times 10^{-5}} M\textsubscript{\(\odot\)}$. The values we obtained for $M_{OI} $ are comparable to the values reported for other novae by \citet{1994ApJ...426..279W}.

\begin{table}
    \centering
    \caption{The mass of neutral oxygen in the ejecta.}
        \begin{tabular}{cccccc}
    \hline \hline
    Day &$\frac{F_{\lambda 6300}}{F_{\lambda 6364}}$ & $\tau_{\lambda6300}$ &$\frac{F_{\lambda 6300}}{F_{\lambda 5577}}$ &$T_e$(K)&  $ M_{OI} (M\textsubscript{\(\odot\)}  \times 10^{-5})$\\
    \hline
       43   &$3.05\pm 0.08$  &0.172 &$2.84\pm 0.08 $  &5276&$1.62$  \\
       111   &$2.8\pm 0.1 $&0.172&$5.4\pm 0.3$&4665&$5.72$\\
       143 & $3.3\pm 0.8 $&0.172&$ 5\pm 1$& 4829&$0.28$\\
       \hline
    \end{tabular}
    
    \label{tab:physical_res}
\end{table}

\subsection{Determination of the electron density}

\begin{figure}[ht]
\includegraphics[width=0.9\textwidth]{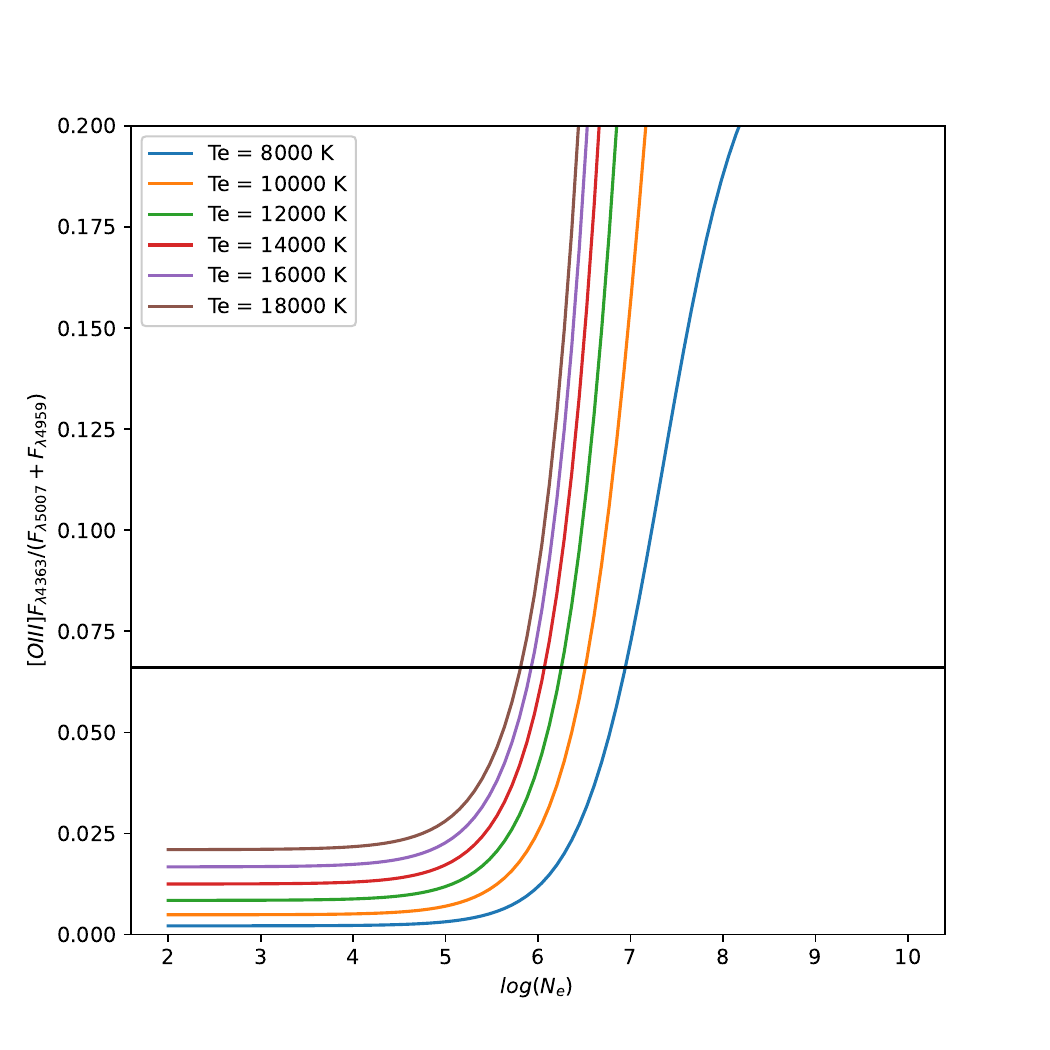}
\caption{$[OIII]F_{\lambda 4363}/(F_{\lambda 4959} + F_{\lambda 5007})$ vs Electron Density for different Temperature Densities generated by PyNeb. Our observed line ratio (the horizontal line) is used to determine the electron density of the ejecta.}
\label{fig:o_ne}
\end{figure}

We used the [O III] lines to determine the $\mathrm{n_e}$ using the ratio of 
\begin{equation}
    \frac{F_{\lambda 4363}}{F_{\lambda 4959} + F_{\lambda 5007}} 
\end{equation}
The ratio yielded a value of $0.066 \pm 0.009$ from the spectra taken on \textbf{Day 515}. We used PyNeb \citep{2015A&A...573A..42L} nebular analysis code assuming several temperatures and densities and then plotting the ratio of the line emissivities (equation 5) as a function of the electron density ($n_e$) computed for different electron temperatures to get Figure \ref{fig:o_ne}. Our ratio yielded initial estimates for the electron densities and electron temperatures which were used to make initial estimates for the abundances of oxygen and nitrogen abundances using PyNeb. These values are listed in Table ~\ref{tab:pyneb}.

\begin{table}
\caption{Initial PyNeb Values.}

\label{tab:pyneb}

\begin{longtable}{lccc}
\hline \hline
$T_e$ (K)   & $log(n_e)$ ($\mathrm{cm^{-3}}$) & O$^a$  &    N$^a$   \\
\hline
8000 &  6.943 	 &  -3.915    &   -4.568      \\
10000 &  6.509	 & -4.695      & -5.245         \\
12000  &  6.252   & -5.166       &  -5.653          \\
14000 &  6.069	 & -5.487      &  -5.934        \\
16000&  5.925	 & -5.723      &  -6.143         \\
18000&  5.810	 & -5.903     &  -6.304         \\
\hline
\end{longtable}
\caption*{$^a$ Log abundance by number relative to Hydrogen.}
\end{table}

\clearpage

\subsection{Photoionization Analysis}

%\begin{figure}[ht]
%\includegraphics[width=0.9\textwidth]{best.pdf}
%\caption{Best Fitting Cloudy Model. }
%\label{fig:synspec}
%\end{figure}

\begin{figure}[ht]
\includegraphics[width=0.9\textwidth]{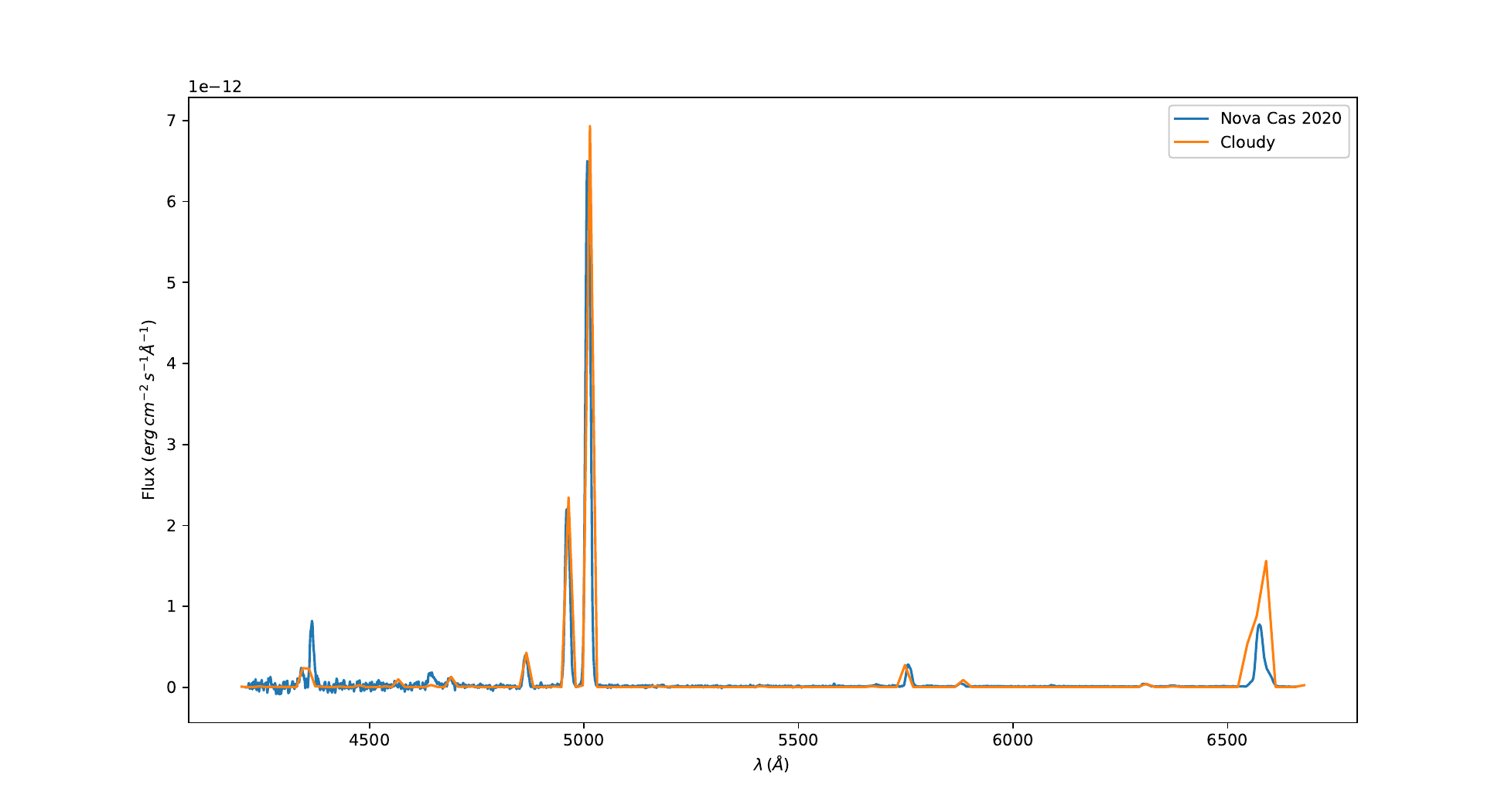}
\caption{Best Fitting Cloudy Model. }
\label{fig:synspec}
\end{figure}

We used the photoionization code Cloudy version C17.02 \citep{2017RMxAA..53..385F} to model the envelope of the nova based on the values we obtained from the nebular analysis of the spectrum on \textbf{Day 515} (Table ~\ref{tab:pyneb}) as initial values (hereinafter referred to as "input"). Starting from these values, we assumed values of log($n_H$) to be equal to the electron densities in Table ~\ref{tab:pyneb}. We ran thousands of models assuming we have a blackbody ionizing source (The underlying WD) with temperatures ranging from 80000 to 200000 K (input 1). We assumed the inner radius of the envelope to be 15.19 (input 2) as an initial estimate based on the average expansion speed determined from the width of the H$\alpha$ line (input 3 as 780 $\mathrm{ km\,s^{-1}}$). We varied the filling factor from 0.05 to 0.8 (input 4). The log of the luminosity of the white dwarf was varied from 35.5 to 36.5 $\mathrm{erg\, s^{-1}}$ (input 5). The parameters of the best-fitting model are tabulated in Table ~\ref{tab:best_fit}.  The synthetic spectrum of this model is plotted in Fig ~\ref{fig:synspec}. The chi-squared goodness of fit test of this model concerning the observed spectrum yielded  $16\%$. Table ~\ref{tab:best_lines} compares the observed emission lines with the lines calculated by the model.

%ab9_5.8_36.5_16.0_0.9_200000.0_0.05

%abund = {'He' : -0.7, 'C' : -5.5, 'N' : -2.0, 'O' : -2.5, 'Mg' :-2.74,'Ne' : -4.00, 'Fe' : -8} ab9

%H	    HELI		CARB	NITR	OXYG	NEON	SODI	ALUM	SILI	IRON
%5.80	5.10		0.30	3.80	3.30	1.80	0.13	0.27	1.34	-2.20

\begin{table}
\centering
\caption{Cloudy Best Fit Parameters.}    
\label{tab:best_fit}
\begin{tabular}{lc}
\hline \hline
Parameter  & Value \\
\hline
log $n_H$ ($\mathrm{cm^{-3}}$)  & 5.81 \\
$T_{BB}$ (K) &   $2\times 10^5$     \\
log luminosity ($\mathrm{erg\, s^{-1}}$) &    36.5     \\
Filling Factor &  0.05     \\
log ($R_{in}$ ) (cm) &   16.0    \\
He $^a$ &  -0.7      \\
C &  -5.5    \\
O &  -2.5    \\
N &  -2.0    \\
Ne  &   -4.00    \\
\hline 
\end{tabular}
\caption*{$^a$ Log abundance by number relative to Hydrogen.}

\end{table}

\begin{table}
\caption{Best Fit Cloudy Emission Lines.}

\label{tab:best_lines}

\begin{longtable}{lcccc}
\hline \hline
 Line ID    & $\lambda$ (\AA)	& Modelled & Observed  &    $\chi^2$   \\
 \hline
H$\gamma$  &  4340 	   & 0.471     &  0.479    &   0.0001      \\
H$\alpha$  &  6562	   & 2.79      & 1.25      &  0.85         \\
H$\beta$   &  4861     & 1.0       & 1.0       &  0.0          \\
{[}N II{]} &  5755	   & 0.74      & 0.85      &  0.017        \\
{[}O III{]}&  4363	   & 0.49      & 1.67      &  2.88         \\
{[}O III{]}&  4959	   & 5.74      & 6.46      &  0.09         \\
{[}O III{]}&  5007	   & 17.13     & 18.75     &  0.15         \\
{[}O I{]}  &  6300 	   & 0.07      & 0.14      &  0.06         \\
{[}O I{]}  &  6364	   & 0.023     & 0.075     &  0.11         \\
{[}O II{]} &  7320	   & 0.077     & 0.256     &  0.415        \\
\hline
Total $\chi^2$     &           &           &           &  4.58         \\
\hline
\end{longtable}
\end{table}

%\begin{table}
%\label{tab:oi_f}
%\begin{longtable}{cccc}
%\hline \hline
%JD (2459000+) & line&Integrated Flux $erg \, cm^{-1} \,s{-1} \, \si{\angstrom}^{-1}$       \\
%169.5321 & N II 5755 & $6.25 \pm 0.07 times 10 ^{-11}$ \\
%\hline
%\end{longtable}
%\end{table}

%\begin{table}
%\label{tab:oi_f}
%\begin{longtable}{ccc}
%\hline \hline
%JD (2459000+) & line&Integrated Flux $erg \, cm^{-1} \,s{-1} \, \si{\angstrom}^{-1}$       \\
%169.5321 & C IV 5805 & $2.76 \pm 0.02 \times 10 ^{-11}$  \\
%\hline
%\end{longtable}
%\end{table}

\subsection{Total Mass of the Ejecta}

We follow the method used by \citet{2006A&A...459..875E} to calculate the total mass of hydrogen in the ejecta using the equation:

\begin{equation}
    M_H = 4 \pi n_e m_H \epsilon \delta R^2
\end{equation}

where $n_e$ is the electron density, $m_H$ is the mass of the hydrogen atom, $\epsilon$ is the filling factor calculated from the flux of the $H_{\alpha}$ line from the equation,
\begin{equation}
    \epsilon = \frac{F_H}{g\,n_e^2\,V}
\end{equation}

where $F_H$ is the flux of $H_{\alpha}$ or $H_{\beta}$ line and $g$ is the volume emissivity of the used line, V is the volume of the envelope given by

\begin{equation}
    V = 4\pi \times R \times \delta
\end{equation}
we get the radius of the envelope at day 515 by using the average expansion velocity $v_{{exp}_{av}} \sim 350 km/s$.

$\delta$ is the thickness of the envelope calculated from the equation:

\begin{equation}
    \delta = R \frac{v_{th}}{v_{exp}}
\end{equation}
where $v_{th}$ is the thermal velocity given by,

\begin{equation}
    v_{thermal}=\sqrt\frac{3kT_e}{m_p}
\end{equation}

where k is the Boltzmann constant, $T_e$ is the electron velocity and $m_p$ is the proton mass \citealp[and references,therein]{2006A&A...459..875E,2019Ap.....62..475T}. 
 We used the values of $T_e = 18000\, K $ and $log(n_e) = 5.81 $ since these values yielded the best-fit cloudy model. We used PyNeb to calculate the emissivities of $H_{\alpha}$ and $H_{\beta}$ lines for the mentioned temperature and densities. we list the calculated physical parameters in Table~\ref{tab:physical_params}.

 \begin{table}[]
     \centering
     \begin{tabular}{l|c}
     \hline \hline
     parameter & value\\
     \hline
      $v_{th}$    & $21\, km\, s^{-1}$\\
       R   & $1\, \times 10^{16} \,cm$\\
       $\delta$ & $6.05\, \times 10^{14}$\\
       V & $7.6\, \times 10^{47} \mathbf{cm^{3}}$\\
       $g_{\alpha}$ & $2.0\, \times 10^{-25}$\\
       $g_{\beta}$ & $7.25\, \times 10^{-26}$\\
       $\epsilon_{{H}_{\alpha}}$ & 0.02  \\
       $\epsilon_{{H}_{\beta}}$ & 0.05\\
       $M_{H}(H_{\alpha})$ & $1.05\, \times 10^{-5}$  $  M\textsubscript{\(\odot\)}$\\
       $M_{H}(H_{\beta})$ & $2.3\, \times 10^{-5}$ $ M\textsubscript{\(\odot\)}$\\
       $M_{tot}(H_{\alpha})$ & $2.1\, \times 10^{-5}$  $  M\textsubscript{\(\odot\)}$\\
       $M_{tot}(H_{\beta})$ & $4.6\, \times 10^{-5}$ $ M\textsubscript{\(\odot\)}$\\
       \hline
     \end{tabular}
     \caption{Physical parameters of the nova envelope on day 515}
     \label{tab:physical_params}
 \end{table}

Our calculated values for the mass of hydrogen range between  $1.05\, \times 10^{-5}$  $  M\textsubscript{\(\odot\)}$ and  $2.3\, \times 10^{-5}$ $ M\textsubscript{\(\odot\)}$. Our abundance analysis suggests that hydrogen comprises about 0.5 of the mass of the envelope which yields a total mass of the ejected envelope ($M_{tot}$) ranging between  $2.1\, \times 10^{-5}$  $  M\textsubscript{\(\odot\)}$ and $4.6\, \times 10^{-5}$ $ M\textsubscript{\(\odot\)}$ where 
\begin{equation}
M_{tot} = \frac{M_H}{X}     
\end{equation}
 where $X$ is the hydrogen mass fraction. Our values are consistent with the masses of most of the novae (see \citealp{2019Ap.....62..475T,2020A&ARv..28....3D}).
 
\section{Discussion and Summary}

The spectrum of the nova evolved from being optically thick dominated by absorption lines to optically thin spectra in less than 30 days. In such an early phase, the outer parts of the ejecta became exposed to emission from hotter regions enhancing the ionization which is shown in the increasing flux of the emission components of the lines and the disappearance of the absorption components. The agreement between the observed flux ratio of the [O I] $\lambda\lambda$ 6300 $\si{\angstrom}$, 6364 $\si{\angstrom}$  doublet (${F_{\lambda 6300}}/{F_{\lambda 6364}}$), and the theoretical value could be attributed to that the ejected envelope of V1391 Cas is optically thin and it is devoid of the dense globules responsible for the [O I] optical thickness in other novae. The absorption components of the Hydrogen lines disappeared by \textbf{Day 43}. The emission lines got stronger and the spectra evolved towards the nebular phase. The absence of forbidden lines of high ionization species in \textbf{Day 43} spectra suggests that the nova was in the pre-nebular phase. The nova transitioned to the nebular phase before \textbf{Day 111} since we see the strongest emission lines where the ejecta is optically thin and exposed to radiation from the hot white dwarf. Our last observation on \textbf{Day 515}, shows multiple [O III] lines along with the [N II] 5577 $\si{\angstrom}$ which shows that the nova is in the late nebular stage.

\clearpage

\begin{figure}
\centering
    \begin{subfigure}[t]{0.45\textwidth}
        \centering
        \includegraphics[width=0.86\linewidth]{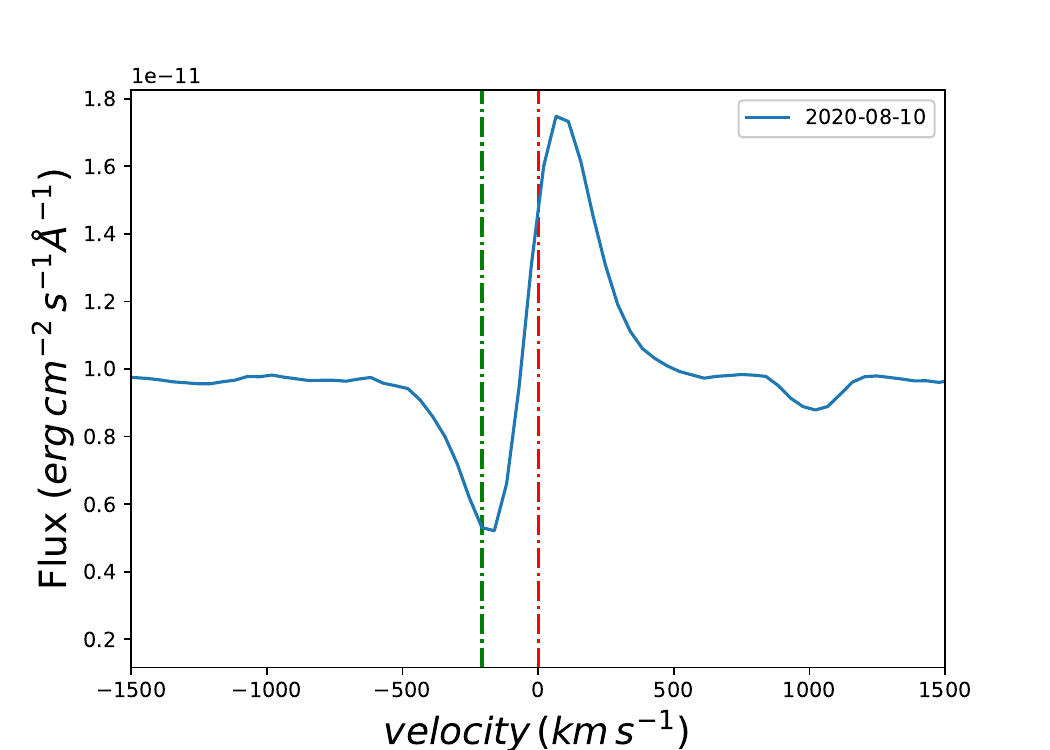} 
        \caption{H$\alpha$.} \label{fig:halpha1}

    \end{subfigure}
    \hfill
    \begin{subfigure}[t]{0.45\textwidth}
        \centering
        \includegraphics[width=0.86\linewidth]{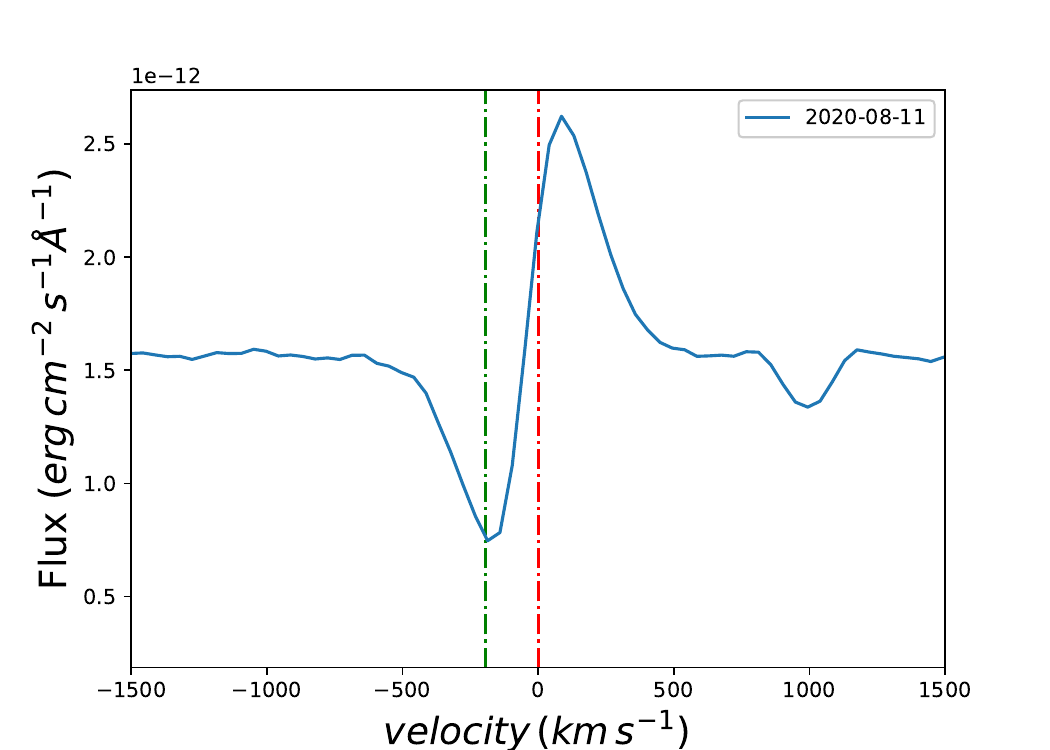} 
        \caption{Same as Fig. \ref{fig:halpha1} for the following night.} \label{fig:halpha2}
    \end{subfigure}
    \vspace{1cm}
    \begin{subfigure}[t]{0.45\textwidth}
    \centering
        \includegraphics[width=0.86\linewidth]{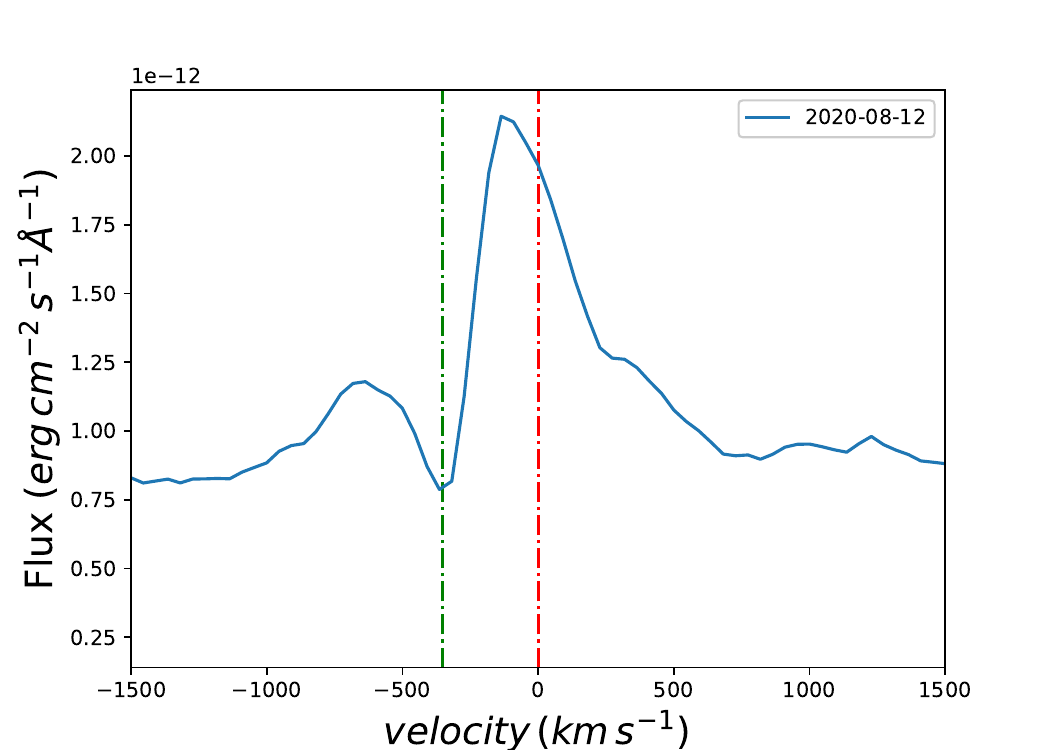} 
        \caption{Same as Fig. \ref{fig:halpha1} for the third night. However here the green line represents the fast mass ejection component.} \label{fig:halpha3}
    \end{subfigure}
    \hfill
        \begin{subfigure}[t]{0.45\textwidth}
        \centering
        \includegraphics[width=0.86\linewidth]{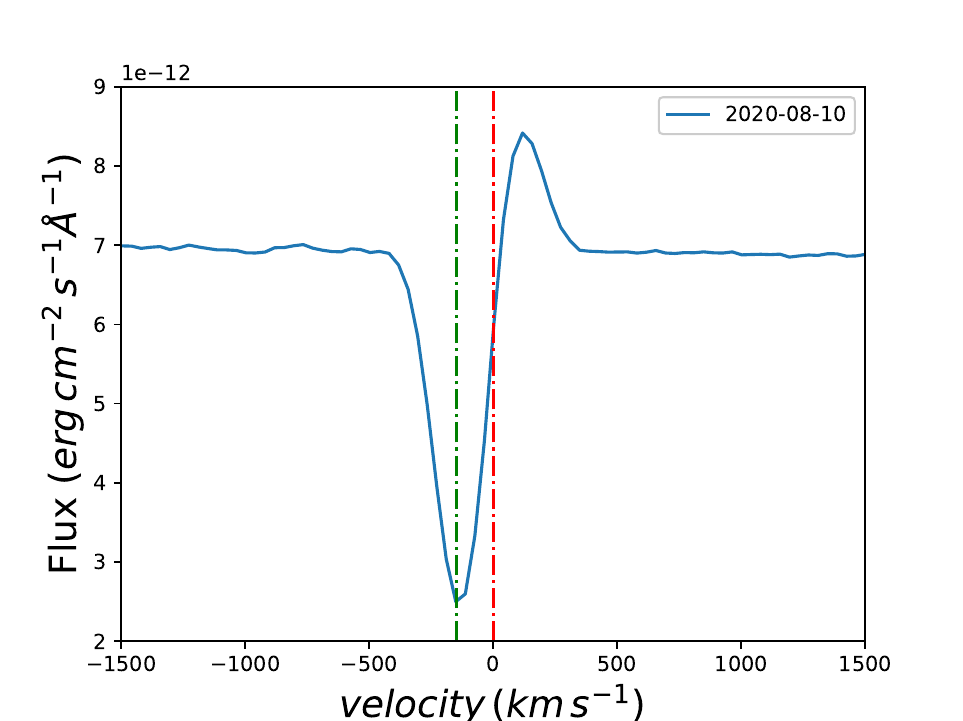} 
        \caption{O I 7772 $\si{\angstrom}$. } \label{fig:oi1}
    \end{subfigure}
    \begin{subfigure}[t]{0.45\textwidth}
        \centering
        \includegraphics[width=0.86\linewidth]{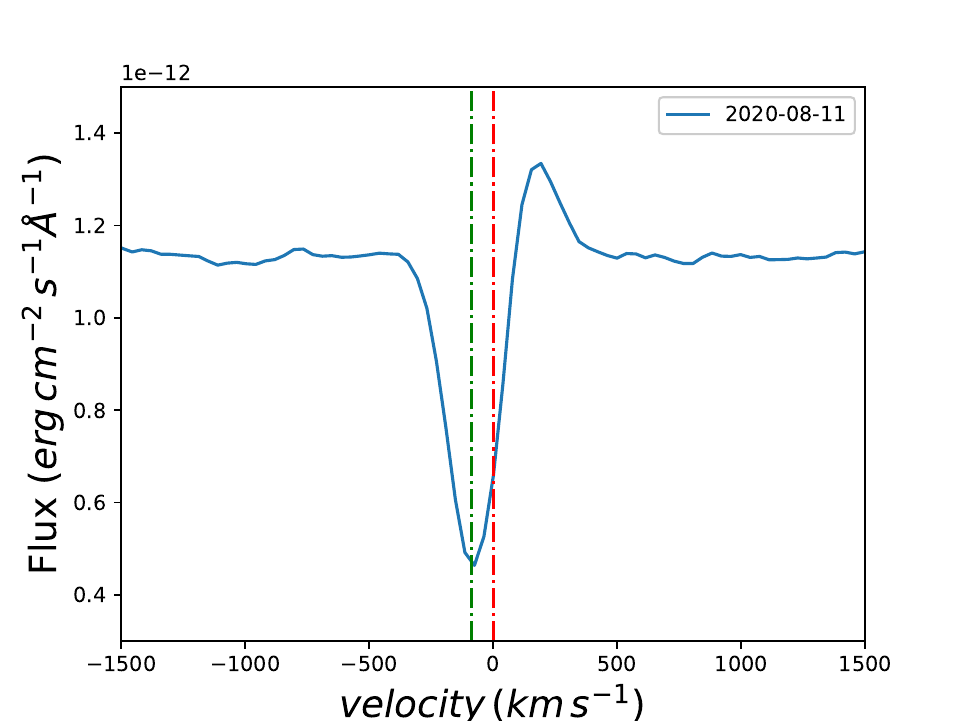} 
        \caption{Same as Fig. \ref{fig:oi1} for the following night.} \label{fig:oi2}
    \end{subfigure}
    \begin{subfigure}[t]{0.45\textwidth}
        \centering
        \includegraphics[width=0.86\linewidth]{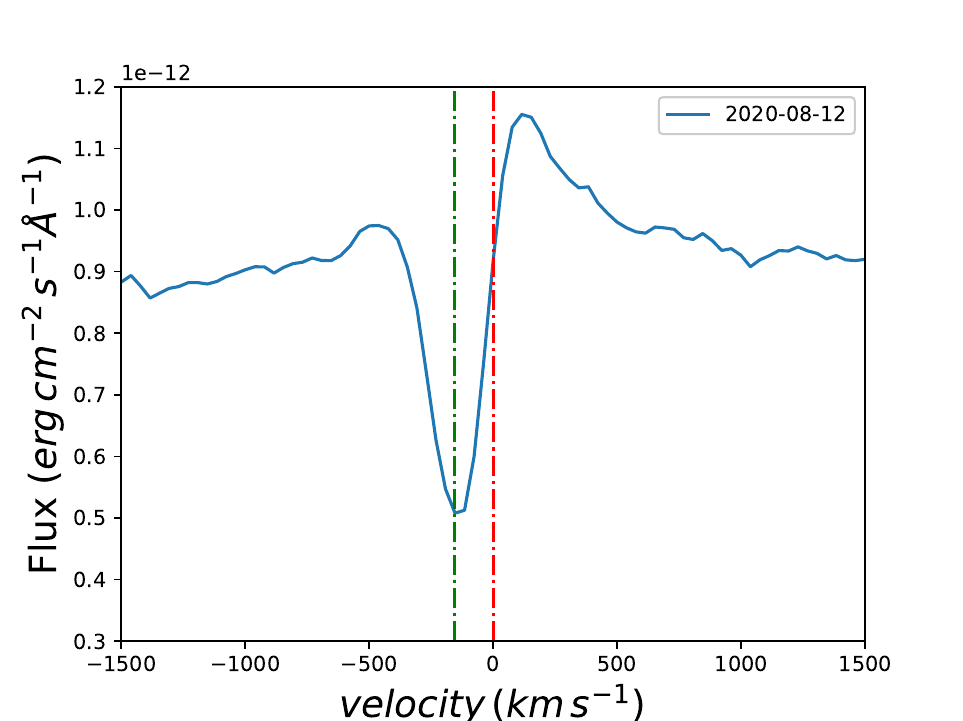} 
        \caption{Same as Fig. \ref{fig:oi1} for the third night.} \label{fig:oi3}
    \end{subfigure}
    \caption{The P-Cygni profile of  two prominent lines. The red dash-dotted line represents the rest velocity. The green line represents the slow velocity component.}
\end{figure}

\citet{2020ApJ...905...62A} showed that classical novae can have multiple distinct outflows of mass ejection near the optical peak. Two of these outflows can be seen as a slow component and a fast component of P-Cygni profiles in lines near the optical peak. During the first three nights of observations (near the optical peak),  our spectra showed two of these outflows in the absorption component of the P-Cygni profiles of some lines mainly H$\alpha$ and O I 7772 $\si{\angstrom}$. These components consist of firstly a slow component with speeds of $\sim$ -200 $\mathrm{km\,s^{-1}}$ and -190 $\mathrm{km\,s^{-1}}$ for the H$\alpha$ line in \textbf{Day 14} and \textbf{15} observations (Figs. \ref{fig:halpha1} and \ref{fig:halpha2}) then the fast component of mass ejection appeared at a speed of $\sim$ -350 $\mathrm{km\,s^{-1}}$ (Fig. \ref{fig:halpha3}). For the O I 7772 $\si{\angstrom}$ line, slow component had speeds of $\sim$ -150 $\mathrm{km\,s^{-1}}$ and -90 $\mathrm{km\,s^{-1}}$ for \textbf{Day 14} and \textbf{15} observations (Figs. \ref{fig:oi1} and \ref{fig:oi2}). For the third night, the speed of the absorption component was at a moderate speed of $\sim$ -150 $\mathrm{km\,s^{-1}}$ (Fig. \ref{fig:oi3} ).

%This  The presence of both a slow and fast component of the mass ejection in classical novae near the optical peak has been observed in many novae and discussed in detail by 

%The presence of multiple assymetric peaks of some lines 

The increase in the expansion velocity, measured by the H$\alpha$ line, with time during the pre-nebular and nebular phases, is a result of the outer ejecta being exposed to radiation from the central source.

The formation of the optically thick dust obscured the nova and it is now very faint in the optical region of electromagnetic radiation.

Our calculated value for the neutral oxygen mass ($\mathbf{M_{OI} = 2.54 \times 10^{-5} M\textsubscript{\(\odot\)}}$) are slightly higher than the values obtained for other novae [e.g. $M_{OI} = 2.52  \times 10^{-5} M\textsubscript{\(\odot\)}$ \citet{1994ApJ...426..279W} and $M_{OI} = 5.15  \times 10^{-5} M\textsubscript{\(\odot\)}$ \citet{2010PASP..122..898M}]. %The lack of density sensitive emission lines in our spectra prevented us from calculating the electron density in the ejecta and hence estimating the elemental abundances.

Novae show jitters in their light curve as they decline in brightness until they experience a sudden drop in brightness associated with dust formation in the ejecta when the gas cools down to a temperature of about 1400 K. This dust blocks the light from the underlying photosphere (note our attempted observations on day 162). As the shell expands, the dust dilutes geometrically and the photosphere is visible again at a brightness slightly lower than before the dip. This recovery occurred about a year after the outburst and we obtained spectroscopic observations of the nova on \textbf{Day 515}. In these observations, we can see that the ionization is enhanced further leading to strong emission of lines with high excitation energies (e.g. [O III] lines along with the [N II] 5577 $\si{\angstrom}$).

The Galactic coordinates of the nova are l=$118^{o}.93153$, b=$+03^{o}.62587$ (\footnote{\url{https://simbad.cds.unistra.fr}}). If we assume the distance from DR3 is 5.7 kpc, we can calculate the height of the nova from the center plane of our Galaxy as z=360 pc. Fast novae located at z<100 pc are associated with the disk, while slow novae up to z$\leq1000$ pc are associated with the galactic bulge \citep{1992A&A...266..232D}. Nova Cas 2020 is a type of nova that has a slow light curve decay ($t_3=120$ days, \citealp{2022MNRAS.517.6150S}), and it belongs to the group of bulge novae. %On the other hand, fast and bright disk novae tend to belong to the He/N class, while the slow and faint thick-disk/bulge novae form the bulk of the FeII class \citep{1998ApJ...506..818D}. It has been confirmed that V1391 Cas belongs to the Fe II class.

We will be able to measure the size of the nova shell by using the expansion velocity of $\sim780\pm20 \mathrm{ km\,s^{-1}}$ together with the expansion parallax method by observing with narrow bands of H$\alpha$\ and [O III] in the years following the explosion when the radius of the ejecta is large enough to be measured from ground-based instruments such the RTT150-TFOSC in Antalya or DAG 4-meters telescope in Erzurum, and we will be able to obtain the more precise distance of the nova and its absolute brightness at the outburst.

We were able to model the spectra of \textbf{Day 515} using Cloudy and we found that the ejected envelope developed into a diffuse cloud with log Hydrogen density of 5.81 $\mathrm{cm^{-3}}$ and elemental abundances (log by number relative to Hydrogen)  of He =  -0.7, C = -5.5, O = -2.5, N -2.0, Ne = -4.00.

We summarized the results of the spectral evolution and photoionization analysis of V1391 Cas in the following.
\begin{itemize}
    \item We have determined that the ejection's expansion velocity is approximately 780 $\mathrm{ km\,s^{-1}}$.
    \item We measured the electron temperature ($T_e$) to be 4966 K and determined that the average mass of neutral oxygen in the ejecta was $2.54  \times 10^{-5} M\textsubscript{\(\odot\)}$.
    \item We modeled the envelope ejected by the nova 515 days after its discovery and calculated the elemental abundances relative to hydrogen.
    \item The total mass of the ejected envelope of the nova 515 days after its discovery was $\sim3.3 \times 10^{-5} M\textsubscript{\(\odot\)}$.
\end{itemize}

\section*{Aknowledgments}

We thank TUBITAK for partial support in using RTT150 (Russian-Turkish 1.5-m telescope in Antalya) with project numbers 1628 and 949. GMH also thanks Turkey Scholarships Postdoctoral Research Fellowship for the financial support. GMH acknowledges travel support from Science, Technology \& Innovation Funding Authority (STDF) under grant number 45779 during the final stages of this work.  The work of AIG was carried out at the expense of a grant allocated by the KFU (project 0671-2020-0052 of the state task 075-00216-20-05).
This study was supported by the following project of Istanbul University Scientific Research Projects Unit (BAP):
FBA-2020-36956.
We acknowledge Kottamia Center of Scientific Excellence for Astronomy and Space Science, STDF project No. 5217 for providing computational facilities used in part of this work.
We acknowledge with thanks the variable star observations from the AAVSO International Database contributed by observers worldwide and used in this research.
We also acknowledge the British Astronomical Association, Variable Star Section and Svensk Amator Astronomisk Forening, variabelsektionen (Sweden) for the photometric data used in this paper.
We would like to thank Professor \c{S}\"{o}len Balman for proof reading the manuscript and for her important remarks.
We are grateful to the anonymous referees for the helpful suggestions and comments which improved the quality of the manuscript. GMH gratefully thanks Istanbul University, Department of Astronomy and Space Sciences for hospitalities.
 
\bibliographystyle{apj.bst}

\begin{thebibliography}{}

\bibitem[\protect\citeauthoryear{{Astropy Collaboration} et~al.}{{Astropy Collaboration} et~al.}{2018}]{astropy:2018}
{Astropy Collaboration}, et~al. 2018, aj, 156, 123

\bibitem[\protect\citeauthoryear{{Astropy Collaboration} et~al.}{{Astropy Collaboration} et~al.}{2013}]{astropy:2013}
{Astropy Collaboration}, et~al. 2013, \aap, 558, A33

\bibitem[\protect\citeauthoryear{{Aydi} et~al.}{{Aydi} et~al.}{2020}]{2020ApJ...905...62A}
{Aydi}, E., et~al. 2020, \apj, 905, 62

\bibitem[\protect\citeauthoryear{{Banerjee} et~al.}{{Banerjee} et~al.}{2020a}]{2020ATel14272....1B}
{Banerjee}, D.~P.~K., {Anupama}, G.~C., {Munari}, U., {Ghosh}, A., {Omar}, A.,  \& {DOT Team}. 2020a, The Astronomer's Telegram, 14272, 1

\bibitem[\protect\citeauthoryear{{Banerjee} et~al.}{{Banerjee} et~al.}{2020b}]{2020ATel14006....1B}
{Banerjee}, D. P.~K., {Shahbandeh}, M., {Evans}, A.,  \& {Hsiao}, E. 2020b, The Astronomer's Telegram, 14006, 1

\bibitem[\protect\citeauthoryear{Bradley et~al.}{Bradley et~al.}{2020}]{larry_bradley_2020_4044744}
Bradley, L., et~al. 2020, astropy/photutils: 1.0.0

\bibitem[\protect\citeauthoryear{{Cayrel}}{{Cayrel}}{1988}]{1988IAUS..132..345C}
{Cayrel}, R. 1988, in The Impact of Very High S/N Spectroscopy on Stellar Physics, ed. G.~{Cayrel de Strobel} \& M.~{Spite}, Vol. 132, 345

\bibitem[\protect\citeauthoryear{Craig et~al.}{Craig et~al.}{2017}]{matt_craig_2017_1069648}
Craig, M., et~al. 2017, astropy/ccdproc: v1.3.0.post1

\bibitem[\protect\citeauthoryear{{Della Valle} et~al.}{{Della Valle} et~al.}{1992}]{1992A&A...266..232D}
{Della Valle}, M., {Bianchini}, A., {Livio}, M.,  \& {Orio}, M. 1992, \aap, 266, 232

\bibitem[\protect\citeauthoryear{{Della Valle} \& {Izzo}}{{Della Valle} \& {Izzo}}{2020}]{2020A&ARv..28....3D}
{Della Valle}, M.,  \& {Izzo}, L. 2020, \aapr, 28, 3

\bibitem[\protect\citeauthoryear{{Dubovsky}, {Medulka}, \& {Kudzej}}{{Dubovsky} et~al.}{2021}]{2021OEJV..220...37D}
{Dubovsky}, P.~A., {Medulka}, T.,  \& {Kudzej}, I. 2021, in Open European Journal on Varıable Stars, Vol. 220, Proceedings of the 52nd Conference on Variable Stars Research, ed. R.~{Kocian}, 37

\bibitem[\protect\citeauthoryear{{Ederoclite} et~al.}{{Ederoclite} et~al.}{2006}]{2006A&A...459..875E}
{Ederoclite}, A., et~al. 2006, \aap, 459, 875

\bibitem[\protect\citeauthoryear{{Esenoglu}}{{Esenoglu}}{1997}]{1997PASP..109.1285E}
{Esenoglu}, H. 1997, \pasp, 109, 1285

\bibitem[\protect\citeauthoryear{{Evans} et~al.}{{Evans} et~al.}{2005}]{2005MNRAS.360.1483E}
{Evans}, A., {Tyne}, V.~H., {Smith}, O., {Geballe}, T.~R., {Rawlings}, J.~M.~C.,  \& {Eyres}, S.~P.~S. 2005, \mnras, 360, 1483

\bibitem[\protect\citeauthoryear{{Ferland} et~al.}{{Ferland} et~al.}{2017}]{2017RMxAA..53..385F}
{Ferland}, G.~J., et~al. 2017, \rmxaa, 53, 385

\bibitem[\protect\citeauthoryear{{Fitzpatrick} et~al.}{{Fitzpatrick} et~al.}{2019}]{2019ApJ...886..108F}
{Fitzpatrick}, E.~L., {Massa}, D., {Gordon}, K.~D., {Bohlin}, R.,  \& {Clayton}, G.~C. 2019, \apj, 886, 108

\bibitem[\protect\citeauthoryear{{Foight} et~al.}{{Foight} et~al.}{2016}]{2016ApJ...826...66F}
{Foight}, D.~R., {G{\"u}ver}, T., {{\"O}zel}, F.,  \& {Slane}, P.~O. 2016, \apj, 826, 66

\bibitem[\protect\citeauthoryear{{Fujii}, {Arai}, \& {Kawakita}}{{Fujii} et~al.}{2020}]{2020ATel13941....1F}
{Fujii}, M., {Arai}, A.,  \& {Kawakita}, H. 2020, The Astronomer's Telegram, 13941, 1

\bibitem[\protect\citeauthoryear{{Fujii}, {Arai}, \& {Kawakita}}{{Fujii} et~al.}{2021}]{2021ApJ...907...70F}
{Fujii}, M., {Arai}, A.,  \& {Kawakita}, H. 2021, \apj, 907, 70

\bibitem[\protect\citeauthoryear{{Hamed}, {Esenoglu}, \& {Galeev}}{{Hamed} et~al.}{2020}]{2020ATel13998....1H}
{Hamed}, G.~M., {Esenoglu}, H.~H.,  \& {Galeev}, A.~I. 2020, The Astronomer's Telegram, 13998, 1

\bibitem[\protect\citeauthoryear{{Jack} et~al.}{{Jack} et~al.}{2020}]{2020AN....341..781J}
{Jack}, D., {Schr{\"o}der}, K.-P., {Eenens}, P., {Wolter}, U., {Gonz{\'a}lez-P{\'e}rez}, J.~N., {Schmitt}, J. H.~M.~M.,  \& {Hauschildt}, P.~H. 2020, Astronomische Nachrichten, 341, 781

\bibitem[\protect\citeauthoryear{{Kalberla} et~al.}{{Kalberla} et~al.}{2005}]{2005A&A...440..775K}
{Kalberla}, P.~M.~W., {Burton}, W.~B., {Hartmann}, D., {Arnal}, E.~M., {Bajaja}, E., {Morras}, R.,  \& {P{\"o}ppel}, W.~G.~L. 2005, \aap, 440, 775

\bibitem[\protect\citeauthoryear{{Kawakita} \& {Arai}}{{Kawakita} \& {Arai}}{2017}]{2017gacv.workE..64K}
{Kawakita}, H.,  \& {Arai}, A. 2017, in The Golden Age of Cataclysmic Variables and Related Objects IV, 64

\bibitem[\protect\citeauthoryear{{Khamitov} et~al.}{{Khamitov} et~al.}{2020}]{2020AstL...46....1K}
{Khamitov}, I.~M., {Bikmaev}, I.~F., {Burenin}, R.~A., {Glushkov}, M.~V., {Melnikov}, S.~S.,  \& {Lyapin}, A.~R. 2020, Astronomy Letters, 46, 1

\bibitem[\protect\citeauthoryear{{Kolb}}{{Kolb}}{2020}]{2020BAVSR..69..193K}
{Kolb}, M. 2020, BAV Rundbrief - Mitteilungsblatt der Berliner Arbeits-gemeinschaft fuer Veraenderliche Sterne, 69, 193

\bibitem[\protect\citeauthoryear{Kramida et~al.}{Kramida et~al.}{2020}]{NIST_ASD}
Kramida, A., {Yu.~Ralchenko}, Reader, J.,  \& {and NIST ASD Team}. 2020, {NIST Atomic Spectra Database (ver. 5.8), [Online]. Available: {\tt{https://physics.nist.gov/asd}} [2020, November 3]. National Institute of Standards and Technology, Gaithersburg, MD.}

\bibitem[\protect\citeauthoryear{{Lenz} \& {Ayres}}{{Lenz} \& {Ayres}}{1992}]{1992PASP..104.1104L}
{Lenz}, D.~D.,  \& {Ayres}, T.~R. 1992, \pasp, 104, 1104

\bibitem[\protect\citeauthoryear{{Luridiana}, {Morisset}, \& {Shaw}}{{Luridiana} et~al.}{2015}]{2015A&A...573A..42L}
{Luridiana}, V., {Morisset}, C.,  \& {Shaw}, R.~A. 2015, \aap, 573, A42

\bibitem[\protect\citeauthoryear{McCully et~al.}{McCully et~al.}{2018}]{curtis_mccully_2018_1482019}
McCully, C., et~al. 2018, astropy/astroscrappy: v1.0.5 zenodo release

\bibitem[\protect\citeauthoryear{{McLaughlin}}{{McLaughlin}}{1965}]{1965POMic...9..113M}
{McLaughlin}, D.~B. 1965, Publications of Michigan Observatory, 9, 113

\bibitem[\protect\citeauthoryear{{Munari} et~al.}{{Munari} et~al.}{2020a}]{2020ATel14267....1M}
{Munari}, U., {Banerjee}, D. P.~K., {Castellani}, F., {Dallaporta}, S., {Maitan}, A.,  \& {Vagnozzi}, A. 2020a, The Astronomer's Telegram, 14267, 1

\bibitem[\protect\citeauthoryear{{Munari} et~al.}{{Munari} et~al.}{2010}]{2010PASP..122..898M}
{Munari}, U., {Henden}, A., {Valisa}, P., {Dallaporta}, S.,  \& {Righetti}, G.~L. 2010, \pasp, 122, 898

\bibitem[\protect\citeauthoryear{{Munari} et~al.}{{Munari} et~al.}{2020b}]{2020ATel13905....1M}
{Munari}, U., {Siviero}, A., {Vagnozzi}, A., {Sokolovsky}, K.,  \& {Aydi}, E. 2020b, The Astronomer's Telegram, 13905, 1

\bibitem[\protect\citeauthoryear{{Oke}}{{Oke}}{1990}]{1990AJ.....99.1621O}
{Oke}, J.~B. 1990, \aj, 99, 1621

\bibitem[\protect\citeauthoryear{{Russell} et~al.}{{Russell} et~al.}{2020}]{2020ATel13967....1R}
{Russell}, R.~W., {Sitko}, M.~L., {Rudy}, R.~J., {Fujii}, M., {Arai}, A.,  \& {Kawakita}, H. 2020, The Astronomer's Telegram, 13967, 1

\bibitem[\protect\citeauthoryear{{Schaefer}}{{Schaefer}}{2022}]{2022MNRAS.517.6150S}
{Schaefer}, B.~A. 2022, \mnras, 517, 6150

\bibitem[\protect\citeauthoryear{{Schlafly} \& {Finkbeiner}}{{Schlafly} \& {Finkbeiner}}{2011}]{2011ApJ...737..103S}
{Schlafly}, E.~F.,  \& {Finkbeiner}, D.~P. 2011, \apj, 737, 103

\bibitem[\protect\citeauthoryear{{Schlegel}, {Finkbeiner}, \& {Davis}}{{Schlegel} et~al.}{1998}]{1998ApJ...500..525S}
{Schlegel}, D.~J., {Finkbeiner}, D.~P.,  \& {Davis}, M. 1998, \apj, 500, 525

\bibitem[\protect\citeauthoryear{{Selvelli} \& {Gilmozzi}}{{Selvelli} \& {Gilmozzi}}{2019}]{2019A&A...622A.186S}
{Selvelli}, P.,  \& {Gilmozzi}, R. 2019, \aap, 622, A186

\bibitem[\protect\citeauthoryear{{Shore} et~al.}{{Shore} et~al.}{2020}]{2020ATel13939....1S}
{Shore}, S.~N., {Boussin}, C., {Boyd}, D., {Buil}, C., {Lester}, T., {Michelet}, J.,  \& {Somogyi}, P. 2020, The Astronomer's Telegram, 13939, 1

\bibitem[\protect\citeauthoryear{{Sokolovsky} et~al.}{{Sokolovsky} et~al.}{2020a}]{2020ATel13903....1S}
{Sokolovsky}, K., et~al. 2020a, The Astronomer's Telegram, 13903, 1

\bibitem[\protect\citeauthoryear{{Sokolovsky} et~al.}{{Sokolovsky} et~al.}{2020b}]{2020ATel13919....1S}
{Sokolovsky}, K., et~al. 2020b, The Astronomer's Telegram, 13919, 1

\bibitem[\protect\citeauthoryear{{Sokolovsky} et~al.}{{Sokolovsky} et~al.}{2020c}]{2020ATel14004....1S}
{Sokolovsky}, K.~V., et~al. 2020c, The Astronomer's Telegram, 14004, 1

\bibitem[\protect\citeauthoryear{{Starrfield}, {Iliadis}, \& {Hix}}{{Starrfield} et~al.}{2016}]{2016PASP..128e1001S}
{Starrfield}, S., {Iliadis}, C.,  \& {Hix}, W.~R. 2016, \pasp, 128, 051001

\bibitem[\protect\citeauthoryear{{Tanaka}}{{Tanaka}}{2011}]{2011PASJ..63...911T}
{Tanaka}, J. 2011, Publ. Astron. Soc. Japan, 63, 911

\bibitem[\protect\citeauthoryear{{Tarasova}}{{Tarasova}}{2019}]{2019Ap.....62..475T}
{Tarasova}, T.~N. 2019, Astrophysics, 62, 475

\bibitem[\protect\citeauthoryear{{Tody}}{{Tody}}{1993}]{1993ASPC...52..173T}
{Tody}, D. 1993, in Astronomical Society of the Pacific Conference Series, Vol.~52, Astronomical Data Analysis Software and Systems II, ed. R.~J. {Hanisch}, R.~J.~V. {Brissenden}, \& J.~{Barnes}, 173

\bibitem[\protect\citeauthoryear{{van Dokkum}}{{van Dokkum}}{2001}]{2001PASP..113.1420V}
{van Dokkum}, P.~G. 2001, \pasp, 113, 1420

\bibitem[\protect\citeauthoryear{{Waagen}}{{Waagen}}{2020}]{2020AAN...715....1W}
{Waagen}, E.~O. 2020, AAVSO Alert Notice, 715, 1

\bibitem[\protect\citeauthoryear{{Warner}}{{Warner}}{2003}]{2003cvs..book.....W}
{Warner}, B. 2003, Cataclysmic Variable Stars (Cambridge: Cambridge University Press)

\bibitem[\protect\citeauthoryear{{Wenzel}}{{Wenzel}}{2021}]{2021BAV.....1....1W}
{Wenzel}, K. 2021, BAV Rundbrief, 1, 44

\bibitem[\protect\citeauthoryear{{Williams}}{{Williams}}{1994}]{1994ApJ...426..279W}
{Williams}, R.~E. 1994, \apj, 426, 279

\bibitem[\protect\citeauthoryear{{Woodward}, {Banerjee}, \& {Evans}}{{Woodward} et~al.}{2020}]{2020ATel14034....1W}
{Woodward}, C.~E., {Banerjee}, D. P.~K.,  \& {Evans}, A. 2020, The Astronomer's Telegram, 14034, 1

\end{thebibliography}

\end{document}